\documentclass[11pt]{article}
\pdfoutput=1
\usepackage{jheppub}
\usepackage[T1]{fontenc}
\usepackage{cancel,tensor,enumitem,fix-cm}
\usepackage{epsfig}
\usepackage{graphicx}
\usepackage[english]{babel}
\usepackage{amsmath,amssymb}
\usepackage{dsfont}
\usepackage{textcomp}
\usepackage{multirow}
\usepackage{booktabs}
\usepackage{tikz}
\usepackage{overpic}
\usepackage{hyperref}
\usepackage{bm}
\usepackage{physics}
\usepackage[lofdepth,lotdepth]{subfig}
\usepackage{float}
\usepackage{xcolor}
\usepackage{caption}

\newcommand{\be}{\begin{equation}}
\newcommand{\ee}{\end{equation}}
\newcommand{\bea}{\begin{eqnarray}}
\newcommand{\eea}{\end{eqnarray}}
\newcommand{\mt}[1]{\textrm{\tiny #1}}

\def\({\left(}
\def\){\right)}
\def\[{\left[}
\def\]{\right]}

\def\p{\partial}

\graphicspath{{./img/}}

\subheader{\begin{flushright}
\texttt{UTTG-14-2021, BRX-TH-6691}
\end{flushright}}

\title{Speeding up the spread of quantum information in chaotic systems}

\author[a]{Stefan Eccles,}
\author[a]{Willy Fischler,}
\author[a]{Tyler Guglielmo,}
\author[b,c]{Juan F. Pedraza}
\author[a]{and Sarah Racz}
\affiliation[a]{Theory Group, Department of Physics,
The University of Texas at Austin, Austin TX, USA}
\affiliation[b]{Department of Physics and Astronomy, University College London, London WC1E 6BT, UK}
\affiliation[c]{Martin Fisher School of Physics, Brandeis University, Waltham MA 02453, USA}

\emailAdd{stefan.eccles@utexas.edu}
\emailAdd{fischler@physics.utexas.edu}
\emailAdd{j.pedraza@ucl.ac.uk}
\emailAdd{tylerg@utexas.edu}
\emailAdd{racz.sarah@utexas.edu}

\abstract{We explore the effect of introducing mild nonlocality into otherwise local, chaotic quantum systems, on the rate of information spreading and associated rates of entanglement generation and operator growth.  We consider various forms of nonlocality, both in 1-dimensional spin chain models and in holographic gauge theories, comparing the phenomenology of each.  Generically, increasing the level of nonlocality increases the rate of information spreading, but in lattice models we find instances where these rates are slightly suppressed.}

\begin{document}
\maketitle
\flushbottom

\section{Introduction}
\label{sec:intro}

The dynamics of entanglement generation and the growth of local Heisenberg operators in chaotic quantum systems have been subjects of intense investigation in recent years, bringing together the interests of condensed matter, quantum gravity, and quantum information theory communities \cite{Shenker:2013pqa,Roberts:2014isa,Shenker:2014cwa, Sachdev:2015efa, Maldacena:2015waa, Hosur:2015ylk, Aleiner:2016eni, Kaufman_2016, Gu:2016oyy, Bohrdt:2016vhv, Zhou_2017, Luitz:2017jrn, Nahum:2017yvy, vonKeyserlingk:2017dyr, deBoer:2017xdk, Grozdanov:2017ajz, Leviatan:2017vur, Hallam:2018mmi, White_2018, Xu:2018xfz, You:2018ons, Nahum:2017xiy,Heyl_2018, Dag:2018oym,Zhou:2018myl,Zhou:2019pob,Sahu:2020xdn,Jian:2020qpp,Zanoci:2020jwl,Cheng:2021mop}.  Generation of entanglement is one of the primary features characterizing the thermalization process for out-of-equilibrium states in a wide variety of chaotic systems. Operator spreading, as defined through various operator entanglement measures and through out-of-time-order correlators (OTOCs), has likewise been used to diagnose chaos in an equally diverse set of systems.  In this work we focus on the related notion of information spreading, which is linked to both phenomena.

In any local chaotic quantum system, information that is initially localized will spread, such that its full recovery requires knowledge of an increasing number of degrees of freedom around the location of origin. Suppose we wish to track the spread of information that is initially localized on some test subsystem $T$. The smallest region/subsystem at any time which can be used to fully reconstruct the initial state of $T$ constitutes the ``information cone'' of $T$.  At late times in homogeneous systems, the boundary of this cone grows linearly at a rate dubbed the information velocity, $v_I$  \cite{Couch:2019zni}.  It was argued in \cite{Couch:2019zni} that this velocity is limited both by the rate of entanglement generation and by the speed of local operator spreading.  These disparate phenomena are therefore both brought to bear on a single physical question of how rapidly information delocalizes. This question is itself rather fundamental, inducing a form of causal structure for thermal-scale operators on the theory which may be more stringent than the ultimate relativistic bounds, and which is present even in nonrelativistic quantum systems.

The effect of nonlocality on entanglement growth and on various diagnostics of chaos has been previously explored in holographic theories \cite{Fischler:2018kwt}. Interestingly, introducing nonlocality into the system was found to enhance scrambling as well as the entanglement generation, eluding previous bounds proposed in the context of local quantum field theory. Further studies support this idea, including tantalizing results on the speed up of thermalization, dissipation and complexification rates \cite{Edalati:2012jj,Fischler:2012ff,Couch:2017yil} due to nonlocal effects. In this work, we extend these studies to ask more in detail how the presence of nonlocality affects the rate of information spreading.  The forms of nonlocality we will consider are ``mild,'' in the sense that there is always a finite length scale associated with nonlocal interactions, and we consider regions larger than this length scale. Thus, upon coarse-graining over short distances we can always recover standard local physics. This mild nonlocality is in contrast, for example, to matrix models, or to lattice system with all-to-all couplings which, in the extreme, could render any notion of local subsystems and neighborhoods to be irrelevant.

In this paper we will consider 1-dimensional spin chain systems as well as holographic gauge theories with various forms of mild nonlocality. Even though the presence of all-to-all interactions seems ubiquitous and necessary for fast scrambling \cite{Belyansky:2020bia}, there are examples of systems that exhibit this phenomenon regardless of the absence of the former \cite{Ikeda:2021wfa}. These are spin chains with a specified combination of some sectors, including a subset of next-to-nearest neighbor interactions. This form of mild nonlocality can in fact be realized experimentally. Known examples include materials with high polarizability which, in the presence of a strong electromagnetic field, mimic the presence of such couplings due to the alignment of their molecular dipoles. We believe that the results of this study could thence be important for emerging quantum technologies such as quantum computation and quantum communications. 

We begin the work with a review of related concepts and techniques in section \ref{sec: review}.  We pay particular attention to describe the effective description for information spreading that arises from coarse-graining over short distances, a.k.a. the membrane or hydrodynamic theory, developed in \cite{Jonay:2018yei} and adapted to holographic settings in \cite{Mezei:2018jco}. Section \ref{sec: spin chains}
asks how nonlocality affects information spreading in chaotic, 1-dimensional spin chain systems, and discusses our numerical results for spin chains with next-nearest neighbor interaction and generalizations.  Section \ref{sec: holography} then poses the same questions in holographic systems inspired by the nonlocal spin chains considered previously. These include non-commutative SYM theory (NCSYM) and dipole-deformed SYM theory (DDSYM), both of which include fundamental dipole-type interactions. Even though, NCSYM was already studied in \cite{Fischler:2018kwt}, the methods employed there (shock waves) are technically different to those used in this paper (membrane theory). As expected, the results agree with each other. This provides evidence that the hydrodynamic description is valid for systems with mild nonlocality as the ones considered in this paper. We conclude in section \ref{sec:conclusions} with a comparative discussion of our results and future directions.

\section{Review of concepts and methods}
\label{sec: review}

Before turning to consider the effects of nonlocality, we here summarize the various terms, concepts, and techniques employed in the investigation.  We begin by reviewing two other characteristic quantities, the entanglement velocity $v_E$ and the butterfly velocity $v_B$, and their relationship to the information velocity $v_I$.  We describe our method of numerically tracking the information cone and computing $v_I$ in chaotic 1-D spin chains.  We then provide a review of the membrane theory of entanglement dynamics \cite{Jonay:2018yei,Mezei:2018jco}, and introduce the membrane tension function $\varepsilon(v)$, which encodes information about the aforementioned quantities. We review its origin, use, and calculation, both in spin chains and in holographic contexts.

\subsection{Three characteristic speeds: $v_I$, $v_E$, and $v_B$}
In addition to our primary quantity of interest, the information velocity $v_I$, we will frequently refer to two other velocities characteristic of local quantum chaotic systems: the entanglement velocity $v_E$ and the butterfly velocity $v_B$.  We review these here for future reference.  The former, $v_E$, is not a velocity in the strict sense, but a quantity that controls the rate of entanglement growth in thermalizing states \cite{Hartman:2013qma, Liu:2013iza, Liu:2013qca}.  For large regions and early times (but much larger than local thermal scales) after a homogeneous quench, the entropy of a region $A$ grows linearly according to
\begin{equation}
	S\left[A(t)\right]=s_{\text{th}}~v_E~|\partial A| t + ...\,,
\end{equation}
where $s_{\text{th}}$ is the thermal entropy density, and $|\partial A|$ denotes the area of the boundary of $A$. This linear growth is explained heuristically in terms of an emergent ``entanglement tsunami'', which is valid in the hydrodynamic regime.\footnote{This picture breaks down for small systems \cite{Kundu:2016cgh,Lokhande:2017jik}, however, here we will focus on regions much larger than both the thermal length and the scale introduced by nonlocality.} For strip-like regions, this linear growth regime persists until saturation.  The ellipses include an initial entropy density, as well as sub-extensive contributions which we ignore as subleading. Some dependencies are suppressed in this expression: both $v_E$ and $s_{\text{th}}$ may depend on the energy and charge density (or equivalently, the inverse temperature and chemical potential) of the state.  Additionally, while $v_E$ is usually defined in the context of entropy growth from an unentangled state (or a vacuum state, with $S\left[A(t)\right]$ taken to be the vacuum-subtracted entropy), it can more generally depend on details of the initial state.  For instance, in this work we will consider initial states with a uniform entropy density greater than zero, but less than the thermal value, referring to $f \equiv s_{\text{initial}}/s_{\text{thermal}}$ as the initial entanglement fraction of the state.  In this case, $v_E$ depends on $f$ as well.

The second velocity, $v_B$ \cite{Shenker:2013pqa,Roberts:2014isa}, is associated with the spread of local perturbations, as diagnosed through the growth of local Heisenberg operators.  The thermal expectation value of the squared commutator of one simple local operator and another, displaced in space and time, e.g. $\left<[\mathcal{O}_1(x),U^\dagger(t)\mathcal{O}_2(y)U(t)]^2\right>_\beta$, vanishes as $|x-y|\rightarrow \infty$, but becomes non-negligible over a displacement range that grows with time.  The region over which the expectation value of the square of this commutator is $\mathcal{O}(1)$ serves to define the butterfly cone. After initial transient behavior, the boundaries of such a cone expand linearly at a rate $v_B$. Such cones map the regions of influence of perturbations by local operators.\footnote{Aside from choosing operators that do not represent locally conserved charge, the cone is largely insensitive to the choice of local operators.}

In \cite{Couch:2019zni} it was argued that the information velocity $v_I$ can be understood through an argument that relates it to both $v_E$ and $v_B$.  The argument follows a variation of the Hayden and Preskill protocol of information recovery \cite{Hayden:2007cs}, and states that the information cone of a test site $T$ can be identified, at any given time, by finding the largest region centered on the test site which satisfies two conditions.  First, it must have reached entanglement saturation, and second, it must be within the butterfly cone of the test site $T$.  The former condition indicates that the active degrees of freedom in the region are maximally entangled with degrees of freedom in the complement system.  The latter serves as an indicator that the region is sufficiently scrambled.\footnote{The precise requirement for ``scrambling'' is not rigorously identified.  In the quantum information argument by Hayden and Preskill, the scrambling operation corresponds to a Haar-averaging over unitaries, though this is presumed excessive.  In the variation of \cite{Couch:2019zni}, it is assumed that degrees of freedom within the butterfly cone are sufficiently scrambled (for purposes of the argument), because all local operators at $T$ can be replaced by truncated operators with support only within the (growing) butterfly cone, and within this region the chaotic dynamics is presumed sufficiently close to a ``random'' unitary.}  When both criteria are met by a region $R$, a Hayden-Preskill style protocol for information recovery implies that the complement region $\bar{R}$, and not $R$, can recover the initial data of $T$.\footnote{Actually, the argument implies that the complement region $\bar{R}$, plus any small region from within $R$, $\Delta R$, with Hilbert space dimension at least greater than that of $T$, can recover the information.  In our large region limit we assume $|R|>>|T|$ and treat the difference between the size of $R$ and $R-\Delta R$ as negligible.} Considering larger and larger regions centered on $T$ at a fixed time until one of the two criteria fails identifies the smallest region $R_I$ that can be used to reconstruct the initial data of $T$.  This discussion assumes a sharp transition between regions of recovery, $R$ and $\bar{R}$.  In reality, the ``edge'' of the information cone may be smoothed, however we assume this does not affect the asymptotic scaling of the growth of $R_I$, and therefore $v_I$.

The above discussion alone indicates that either the scrambling condition or the entanglement condition could act as a bottleneck to the spread of information.  However, in all systems considered, the condition of entanglement saturation always sets the more stringent constraint, at least in the large-region, late-time limit.\footnote{This statement is related to several known bounds in the literature.  In \cite{Hartman:2013qma,Casini:2015zua} it was proven that $v_E\le c$ in relativistic systems and then by a causality argument that $t_{\text{sat}}\ge r_{\text{ins}}/c$, where $r_{\text{ins}}$ is the radius of the largest ball that fits in the region being considered. Then in \cite{Mezei:2016wfz, Mezei:2016zxg} it was shown that $v_E\le v_{LC}$ and $t_{\text{sat}}\ge r_{\text{ins}}/v_{LC}$, where $v_{LC}$ is an effective light cone speed, usually assumed equal to $v_B$ (though not always, see \cite{Bohrdt:2016vhv}).} Therefore we henceforth take for granted that $v_I$ can be identified by finding the growth of the largest region that has just reached entanglement saturation. For an arbitrary region $A$, the time to entanglement saturation is not solely controlled by $v_E$, but in cases where linear growth persists until saturation, in the scaling limit it is simply given by $t_{\text{sat}}=(1-f)v_E^{-1}|A|/|\partial A|$. For strips of width $R$ this gives $t_{\text{sat}}=R(1-f)/v_E(f)$. Together with the statement that $v_E(f)\sim v_B(1-f)+...$ near $f=1$ \cite{Mezei:2016wfz, Jonay:2018yei, Couch:2019zni}, this entails that for strip-shaped regions,  $v_I$ interpolates between $v_E$ at $f=0$ and $v_B$ at $f=1$.  Finding $v_I$ for strips across all values of $f$ thereby also provides information about the rates controlling entanglement growth and operator spreading in the system. Computing $v_I(f)$ for strip-like regions will therefore be our primary focus, leaving extensions to other shapes and states to future work.

The fact that $v_I$ is simply related to the largest region to reach entanglement saturation also allows us to relate $v_I$ directly to the membrane tension function, a quantity described in the section \ref{subsec: membrane theory}.  In spin chain systems, we can only roughly approximate the membrane tension function due to computational limitations and finite size effects.  Hence, we are better served by computing $v_I$ directly through a method explained in section \ref{subsec: tracking vI in spin chains}.  Holographic systems, on the other hand, are particularly amenable to the membrane method.  In these systems we compute $v_I$ by relating it to the membrane tension function, as explained in section \ref{subsec: membrane theory in holography}.

\subsection{Tracking the information cone in 1-D spin chains}
\label{subsec: tracking vI in spin chains}

For the purpose of identifying the information cone in a local quantum chaotic system\footnote{Controlled information transport in non-chaotic spin chains is another interesting vein of research.  See, for example, \cite{2007PhRvL..99y0506C} and references therein.}, we imagine that the degrees of freedom on a local test site, $T$, are initially maximally entangled with a reference system $R$ of equal or greater Hilbert space dimension.  The reference $R$ is thereafter completely decoupled from the system dynamics. Initially, $R$ has maximal mutual information with the test degrees of freedom, but as the system evolves, the mutual information between $R$ and $T$ decreases as data from $T$ scrambles into the rest of the system.  The information cone can be identified by tracking the smallest region around $T$ that retains maximal mutual information with the reference system.  Any other initial state at $T$ (besides the maximally mixed state used in this procedure) could, in principle, be inferred from data within this region \cite{Barnum:1998da}. 

For a 1-dimensional chaotic spin chain, we instantiate a version of this setup as follows (this is a slight modification of a procedure employed in \cite{Couch:2019zni}):  We consider $L$-qubit systems with nearest-neighbor $\sigma_z \sigma_z$ couplings, with longitudinal and transverse magnetic field terms chosen to render the Hamiltonian chaotic:
\begin{equation}\begin{split}\label{eq: H local}
		H^{(0)}_L=&J_{zz}\sum_{i=1}^{L-1}\sigma_z^{(i)}\sigma_z^{(i+1)}
		+h_x\sum_{i=1}^{L}\sigma_x^{(i)}
		+h_z\sum_{i=1}^{L}\sigma_z^{(i)}\\
\end{split}\end{equation}
where $J_{zz}=1,~~h_x=1.05,~~ h_z=0.5$.  Variations of this Hamiltonian have been widely used in the literature to study thermalization.  The superscript on $H^{(0)}_L$ merely indicates that this serves as our ``base'' Hamiltonian, to which we will later add various nonlocal couplings.  However, the nonlocal couplings are irrelevant for the demonstrations of this section.
We also define the following set of states:
\begin{equation}\begin{split}\label{eq: f states}
		\ket{\psi_L^f} = \prod_{i=1}^{L/2} a_{\sigma,\sigma'}\ket{\sigma_{i}}\ket{\sigma_{i+L/2}}
\end{split}\end{equation}
with the matrix $a$ given by
\begin{equation}\begin{split}\label{eq: a matrix}
		a=\left(\frac{1}{2}I+\xi ~\sigma_y\right)^{1/2}.
\end{split}\end{equation}
These states allow us to initialize a uniform ``volume law'' entanglement fraction $f$, such that the entanglement entropy of a subsystem of $N_{\text{sub}}<L/2$ neighboring qubits is $f*N_{\text{sub}}$. The $f$ dependence is implicitly related through $\xi$ satisfying $-(1/2+\xi)\log(1/2+\xi)-(1/2-\xi)\log(1/2-\xi)=f \log(2)$.  For all $f$, these states are at the center of the energy spectrum, in that $\expval{H_L}{\psi}=0$.
An additional Bell pair is then appended to the system.   One member of the pair is dynamically coupled to the system while the other remains decoupled.  These are the ``test'' and ``reference'' qubits respectively.
\begin{equation}\begin{split}\label{eq: full initial state}
		\ket{\psi_{L+2}(t=0))} = \left(\ket{0}_{R}\ket{0}_{T}+\ket{1}_{R}\ket{1}_{T}\right)\ket{\psi^f_L}
\end{split}\end{equation}
The system is then evolved with the $L+1$ qubit Hamiltonian \eqref{eq: H local}, treating the test qubit as the first in an $L+1$ qubit spin chain.  The reference and test qubits initially have maximal mutual information, but as the information associated with the test qubit scrambles into the rest of the system, an ever-growing number of nearby qubits are required to recover the mutual information with the reference\footnote{In practice, we use a cutoff value of one bit of mutual information.  The subsystems under consideration, which all include the test qubit and a variable number of its neighbors, evolve from maximal mutual information to near zero mutual information (with the reference) at late times as long as the number of qubits is under half the system size.  A mutual information of 1 captures the center of this falloff, and we use it to define the information ``wavefront.''}. The growth of the smallest subsystem necessary to achieve this defines the growth of the information cone.

\begin{figure}[t!]
\centering
\includegraphics[width=3in]{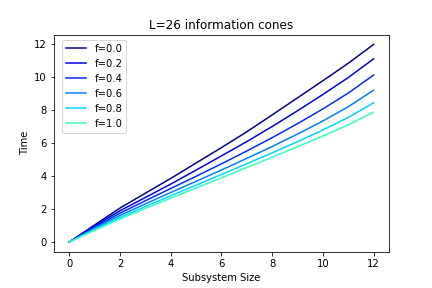}$~$\includegraphics[width=3in]{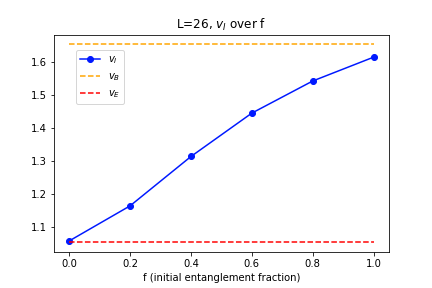}
\begin{picture}(0,0)
\put(-220,70){{\scriptsize $v_I$}}
\end{picture}
\caption{Left: A sample of information cones in the $H^{(0)}_{L=26}$ system for various values of initial entanglement fraction $f$. Right: Information velocity as a function of $f$, extracted from the curves in the left panel.  The dotted lines represent the butterfly velocity $v_B$, as obtained through an independent OTOC computation, and the entanglement velocity $v_E$, computed as the linear growth rate of entanglement on half the system.\label{fig: basic inf cones and vI(f)}}
\end{figure}
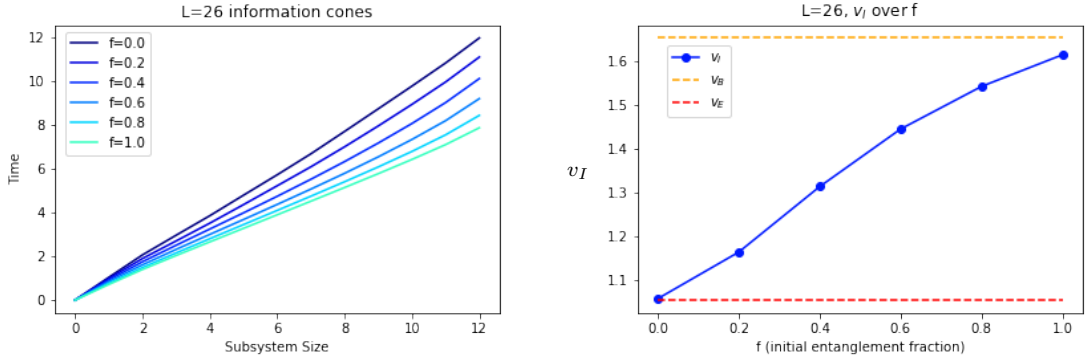

Our Hamiltonian and states have a quasi-translation invariance, voided by the existence of the boundaries, or by the fact that we are computationally restricted to finite $L$.  We therefore track the cone across just under half the system size, extracting $v_I$ as a linear fit on the cone in this region.  Most of our results are for systems with total $L=26$. In fitting $v_I$, we exclude the initial and final three subsystems of the half-system to minimize finite-time and edge effects. We implement time evolution with a Krylov subspace method to exponentiate the Hamiltonian.  This code was written in C, and based on the well known Fortran package, Expokit \cite{EXPOKIT}, where modifications were made to improve efficiency and parallelize the computation.  Each time evolution and entropy computation for a particular Hamiltonian was run on a single node across 44 cores on TACC's Stampede2 computer.  The results were verified for smaller $L$ using the Dynamite \cite{2017_dynamite} and Qutip \cite{qutip1,qutip2} python packages as well as a Mathematica \cite{Mathematica} implementation.  Figure \ref{fig: basic inf cones and vI(f)} shows examples of information cones from states initialized with varying degrees of uniform entanglement, $f$.  The information velocity is plotted as a function of initial entanglement fraction.  As $f\rightarrow 1$, $v_I$ is expected to approach $v_B$.  As a consistency check, we computed $v_B$ independently by tracking the growth of squared commutators of simple operators, finding values in approximate agreement, considering finite size effects in both computations.

\subsection{Membrane theory of entanglement dynamics}
\label{subsec: membrane theory}

An interesting quantity which can serve to simultaneously characterize aspects of entanglement growth, operator dynamics, and information spreading in local quantum chaotic systems is the membrane tension function, ${\cal E}(v)$ \cite{Jonay:2018yei}. The membrane model of entanglement dynamics emerges in a course-grained, hydrodynamic type approximation of such systems.  It computes the entanglement entropy of a state via an integrated energy density along a timelike membrane stretching through an auxiliary Minkowski spacetime, analogous to a minimal cut through a tensor network or quantum circuit representing unitary time evolution. The membrane method has been applied with great success to holographic systems \cite{Mezei:2018jco,Mezei:2019zyt,Agon:2019qgh}, where the membrane is merely a projection of bulk extremal surfaces, with the bulk dimension integrated out.  In this section, we review the origin of the membrane method in a thermodynamic limit of chaotic quantum systems.  We review a method of approximating ${\cal E}(v)$ in 1-D spin chain systems which we will later employ in section \ref{subsec: spin chain approx membranes}.  We then discuss its derivation and use in the context of the AdS/CFT correspondence as well as its extension to more general gauge/gravity dualities.  This section is largely a review of work appearing in \cite{Jonay:2018yei} and \cite{Mezei:2018jco}, though, we generalize the techniques to theories with inherent nonlocality.

\subsubsection{Membrane theory for spin chains}
\label{subsubsec: membrane theory in spin chains}
To motivate the membrane method, we restrict ourselves to states in local chaotic quantum systems exhibiting a ``volume law'' entanglement pattern, such that the entanglement entropy of subregion is simply proportional to the volume of the subregion (plus subleading corrections).  For simplicity we also first consider 1-dimensional systems, though the generalization to higher dimensions is immediate.  In a system of total length $L$, pure states partitioned by a cut at position $x$, at equilibrium exhibit a bipartite entanglement entropy of

\begin{equation}
	S_{\text{eq}}(x,t)=s_{\text{eq}}\min\left(x,L-x\right)
\end{equation}
where $s_{\text{eq}}$ is the equilibrium entanglement density.  The leading, course-grained dynamics of out-of-equilibrium states can be captured in a hydrodynamic description, governed by
\begin{equation}\label{eq: hydrodynamic S}
	\frac{\partial S}{\partial t}= s_{\text{eq}} \Gamma\left(\frac{\partial S}{\partial x}\right).
\end{equation}
Here $\Gamma\left(\frac{\partial S}{\partial x}\right)=\Gamma(s)$ is the entropy production rate.  In general this function could depend explicitly on position or on any locally conserved charge densities, though we consider the simplest, spatially uniform case.  Physical considerations immediately put some constraints on $\Gamma(s)$.  At saturation, entanglement growth vanishes $\frac{\partial S}{\partial t}=0$, implying that $\Gamma(s_{\text{eq}})=\Gamma(-s_{\text{eq}})=0$.  Considering the growth of entanglement from a completely unentangled state leads to $\Gamma(0)=v_E$.  The entropy production rate is only defined over $s_{\text{eq}}\le s \le s_{\text{eq}}$, and we assume a smooth, symmetric function with $\Gamma''(v)<0$ over this domain. 

Next suppose that there is an equivalent representation of the entropy dynamics, staged on an auxiliary Minkowski spacetime associated with the time evolution operator, where the entanglement entropy is computed by minimizing the integrated local ``energy'' along a timelike membrane stretching from the initial to final time and obeying certain boundary conditions.  The tension function ${\cal E}(v)$ specifies the energy density along this membrane as a function of its orientation in the spacetime (its local ``velocity'').  For the one-dimensional translation invariant system described above, the entropy at position $S(x,t)$ would be obtained by the following minimization (see figure \ref{fig: tikz membrane}):

\begin{figure}[h]
	\centering
	\includegraphics[width=15cm]{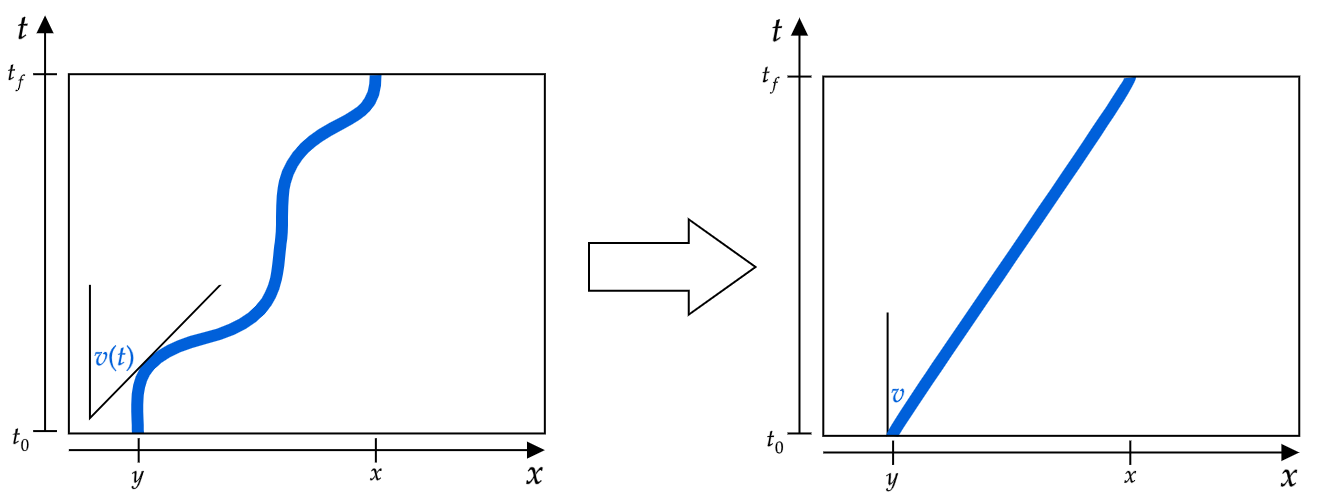}
	\caption{Minimization over membranes brings the curve on the left where a membrane stretches from $y$ to $x$ with a velocity $v(t)$ assigned as the angle from the vertical at each point in time to a straight line.}
	\label{fig: tikz membrane}
\end{figure}

\begin{equation}
	S(x,t)=\underset{y}{\min}\left(t~ s_{\text{eq}}~{\cal E}\left(\frac{x-y}{t}\right)+S(y,0)\right)
\end{equation}

Differentiating with respect to time and comparing to equation \eqref{eq: hydrodynamic S} entails that ${\cal E}(v)$ can be directly related to $\Gamma(s)$ (and vice versa) by Legendre transformation:

\begin{equation}\begin{split}
		\Gamma(s)=
		\underset{v}{\min}\left({\cal E}(v)- \frac{vs}{s_{\text{eq}}}\right)\\
		{\cal E}(v)=
		\underset{s}{\max}\left(\Gamma(s)+\frac{vs}{s_{\text{eq}}}\right)\\
\end{split}\end{equation}

From the properties of $\Gamma(s)$, we can infer related properties of $\varepsilon(v)$.  In particular, considering $\Gamma(s_{\text{eq}})=0$ leads to $v_{\text{max}}=s_{\text{eq}}\Gamma'(s_{\text{eq}})$.  In the tensor network or circuit picture, the maximum velocity $v_\text{max}$ serves as an effective lightcone velocity, which in known cases is equal to the butterfly velocity $v_B$. Results in the systems we consider are consistent with this expectation, and we assume it throughout. In conjunction with the other characteristics of $\Gamma(s)$ and properties of the Legendre transform, in the range $-v_B\le v \le v_B$ we have
\begin{eqnarray}
	{\cal E}(0)=v_{E},&~~ {\cal E}(v_{B})=v_{B},\nonumber \\
	{\cal E}'(0)=0, &~~ {\cal E}'(v_B)=1, \nonumber \\
	{\cal E}(v)\ge |v|, &~~  {\cal E}''(v)\ge 0.
\end{eqnarray}

The definition of $\varepsilon(v)$ can be extended outside this domain, but this is the region of dynamical interest. It is evident that the membrane tension function encodes a lot of information characterizing the system, including not only entanglement dynamics ($v_E$), but also operator growth ($v_B$) and indirectly, information spreading ($v_I$).  For this reason it will be one of the primary quantities which we compute and compare for systems with variable degrees of nonlocality. 

In general it is not possible to compute ${\cal E}(v)$ analytically for a given Hamiltonian.  A numerical approximation technique was introduced in \cite{Jonay:2018yei}, which we summarize here and later employ.  The membrane tension function at infinite temperature can be simply related to the operator entanglement of the time evolution unitary.  Operator entanglement is computed by treating the operator as a state vector in a doubled Hilbert space.  For instance, from the time evolution operator $U(t)$ we obtain:

\begin{equation*}
\bra{i}U(t)\ket{j}\ket{i}\bra{j}
\longrightarrow
U_{ij}(t)\ket{j}\ket{i}
\end{equation*}
Here, the first and second Hilbert space factors can be thought of as representing the same degrees of freedom at time $0$ and time $t$, respectively.  An ordinary entanglement entropy can then be computed on various subsystems of this state.  In a 1-D system, tracing out degrees of freedom up to $x$ on the first factor and up to $y$ on the second factor gives the entanglement entropy $S_U(x,y,t)$, which in the membrane picture corresponds to a minimal membrane stretching from $x$ to $y$ through an auxiliary spacetime of height $t$.  This leads to a method of approximating the membrane tension function via operator entanglements of the time evolution unitary:

\begin{equation*}
{\cal E}_{\text{eff}}\left(t,v\right)\equiv\frac{1}{s_{\text{eq}}t}
S_U\left(-\frac{vt}{2},\frac{vt}{2},t\right)
\end{equation*}
Here, for system of finite length $L$, the origin of the spatial arguments is placed at the center.  At any given $t$, evaluating the right hand side gives an approximation of the membrane tension function at infinite temperature. This approximation is limited by both finite size effects and finite time effects.  In an infinite system, taking a large time limit would render finite time effects negligible, and the approximation would converge to ${\cal E}(v)$ over the range $-v_B<v<v_B$, and to $|v|$ outside this range.  But for a finite system as $t$ increases, the range of $v$ over which the cut can be made without exceeding the boundaries (or more stringently, before edge effects become substantial) becomes more limited. In our spin chain system, we are computationally limited to $13$ qubits.  So in practice, we obtain a series of curves for a set of $t$ values (or rather, discreet data points such that $|vt/2|$ correspond to an integer number of spins traced out) and schematically piece these together to give a qualitative picture of ${\cal E}(v)$ under different Hamiltonians.  This method is employed in section \ref{subsec: spin chain approx membranes} as a qualitative check on our more direct methods of investigating the effect of variable nonlocality in spin chain systems.

\subsubsection{Membrane theory for holographic systems}
\label{subsec: membrane theory in holography}

The membrane theory of entanglement dynamics has been successfully generalized to strongly-coupled large-$N$ theories with holographic duals  \cite{Mezei:2018jco,Mezei:2019zyt}. This provides strong evidence that this effective description can efficiently encapsulate the entanglement dynamics in all chaotic systems. In \cite{Mezei:2018jco} it was first realized that a membrane description naturally emerges upon studying the late-time and large-distance regime of the RT prescription, akin to hydrodynamics. This study was later complemented in \cite{Mezei:2019zyt}
showing that this theory is robust under a large set of generalizations, including different quench protocols and finite coupling corrections holographically dual to higher derivative gravity corrections. In this section we will first review the results of \cite{Mezei:2018jco}, valid for asymptotically AdS spaces in Einstein gravity. We will then generalize this framework to more general holographic scenarios, including those described by non-AdS spaces with extra matter fields and fluxes, ubiquitous in top-down constructions emerging from string or M theory. 

\subsubsection*{The case of asymptotically AdS spaces}

We start with a generic asymptotically AdS$_{d+1}$ black brane
\be\label{metric}
ds^2= {L^2\over z^2}\left[-a(z)dv^2-{2\over b(z)}\,dvdz+ d\vec{x}_{d-1}^2\right]\,,
\ee
where $v$ is an ``infalling'' time coordinate. For instance, for the standard Schwarzschild-AdS$_{d+1}$ we have
\be\label{Schw}
a(z)=1-\left(\frac{z}{z_h}\right)^d\,,\qquad b(z)=1\,,
\ee
with
\be
T=\frac{1}{\beta}=-\frac{1}{4\pi}\frac{d}{dz}a(z)\Big|_{z_h}=\frac{d}{4\pi z_h}\,.
\ee
We follow \cite{Mezei:2018jco} but keep working in cartesian coordinates. In this way, we can later generalize the analysis to non-AdS settings in a more straightforward way. First,  consider the scalings
\be
v\equiv \Lambda\,\tau\,, \qquad x_i \equiv \Lambda \, \xi_i \,,\qquad z\equiv \zeta\,.
\ee
Defining $\xi\equiv\xi_1$, $\vec{\xi}_\perp\equiv\{\xi_2,\ldots,\xi_{d-1}\}$ and parameterizing the RT surface with functions $\zeta(\tau,\vec{\xi}_\perp)$,  $\xi(\tau,\vec{\xi}_\perp)$, one finds that in the limit $\beta/\Lambda\to 0$ the entanglement entropy functional reduces to:
\bea
S_A=\frac{L^{d-1}\Lambda^{d-1}}{4G_N}  \int d\tau d\vec{\xi}_\perp\ {\sqrt{Q}\over \zeta^{d-1}}\,,\qquad Q\equiv (\p_\tau \xi)^2- a(\zeta)\left[1+(\p_{\perp} \xi)^2\right]\,,
\eea
where $(\p_\perp\xi)^2\equiv (\p_{\xi_2}\xi)^2+\cdots+(\p_{\xi_{d-1}}\xi)^2$. In this limit, the equation of motion for $\zeta$ is algebraic:
\be\label{EOMzeta}
{(\p_\tau \xi)^2\over 1+(\p_\perp \xi)^2}=a(\zeta)-{\zeta a'(\zeta)\over 2(d-1)} \,.
\ee
We now define:
\be\label{vcdefs}
v^2\equiv {(\p_\tau \xi)^2\over 1+(\p_\perp \xi)^2}\,,\qquad c(\zeta)\equiv  a(\zeta)-{\zeta a'(\zeta)\over 2(d-1)}\,.
\ee
so that (\ref{EOMzeta}) becomes $v^2=c(\zeta)$. Plugging back into the action the solution to this equation, $\zeta=c^{-1}(v^2)$, we can rewrite the entropy functional as
a membrane functional:
\bea
S_A=\frac{L^{d-1}\Lambda^{d-1}}{4G_N} \int d\tau d\vec{\xi}_\perp\ \sqrt{1+(\p_\perp \xi)^2}\,{\cal E}\left(v\right)
=s_{\text{th}}\Lambda^{d-1} \int d\text{area}\ { {\cal E}\left(v\right)\over \sqrt{1-v^2}}\,,
\eea
where
\be
s_{\text{th}}=\frac{L^{d-1}}{4G_N \zeta_h^{d-1}},
\ee
is the thermal entropy density of the black brane.
In the above we have identified the standard ``area'' element in Minkowski space
\be
d\text{area}=d\tau d\vec{\xi}_\perp\sqrt{1+(\p_\perp \xi)^2-(\p_\tau \xi)^2}
\ee
and defined the membrane tension ${\cal E}(v)$ to be
\be\label{tension}
 {\cal E}\left(v\right)\equiv \zeta_h^{d-1} {\sqrt{- a'(\zeta)\over 2(d-1)\zeta^{2d-3}}}\Bigg\vert_{\zeta=c^{-1}(v^2)}\,.
\ee
Notice that the factor $\zeta_h^{d-1}$ is included in (\ref{tension}) to make ${\cal E}(v)$ dimensionless. For example, for Schwarzschild-AdS$_{d+1}$ (\ref{Schw}) we find that the equation (\ref{EOMzeta}) becomes
\be
v^2=1-\left(\frac{\zeta}{\zeta_h}\right)^d+{d\over 2(d-1)}\left(\frac{\zeta}{\zeta_h}\right)^d=1-\frac{(d-2)}{2(d-1)}\left(\frac{\zeta}{\zeta_h}\right)^d\,,
\ee
which can be inverted as follows:
\be
\zeta=c^{-1}(v^2)=\zeta_h\left[\frac{2(d-1)}{(d-2)}(1-v^2)\right]^{1/d}\,.
\ee
Plugging this back into (\ref{tension}) we find that
\be
 {\cal E}\left(v\right)={v_E\over (1-v^2)^{(d-2)/(2d)}}\,,\qquad v_E={\left(d-2\over d\right)^{(d-2)/(2d)}\over\left(2(d-1)\over d\right)^{(d-1)/d}}\,.
\ee
It can be checked that this function has all the desired properties \cite{Mezei:2018jco,Mezei:2019zyt} (in the dynamically relevant regime, i.e., $0\leq v\leq v_B$):
\be\label{eqsmembrane}
 {\cal E}(0)=v_E\,, \qquad {\cal E}'(0)=0\,, \qquad {\cal E}(v_B)=v_B\,, \qquad {\cal E}'(v_B)=1\,,\qquad{\cal E}'(v)\geq0\,, \qquad {\cal E}''(v)\geq0\,,
\ee
with
\be
v_B=\sqrt{\frac{d}{2(d-1)}}\,.
\ee
One final remark is in order. In general, solving for $v_B$ from (\ref{eqsmembrane}) may be very difficult. However, it is useful to notice that
\be\label{vBeasy}
v_B^2=c(\zeta_h)\,,
\ee
so $c^{-1}(v_B^2)=\zeta_h$ must yield the horizon. On the other hand,
\be\label{vEeasy}
0=c(\zeta_\text{HM})\,,
\ee
and $c^{-1}(0)=\zeta_{\text{HM}}$ yields the Hartman-Maldacena surface \cite{Hartman:2013qma}.

\subsubsection*{General theory for non-AdS spaces}

The gravity duals of the theories we want to study are non-asymptotically AdS and possibly anisotropic. In particular, their metrics are not of the form (\ref{metric}). This means we need to generalize the membrane theory starting from a more general ansatz. For concreteness, we will take the metric to be of the following form:
\be\label{metric2}
ds^2= -g_{vv}(z)dv^2-{2g_{vz}(z)}\,dvdz+ \sum_{i=1}^{d-1}g_{ii}(z)dx_i^2+d\Omega_p^2\,,
\ee
where $d$ is the number of spacetime dimensions in the boundary and $d\Omega_p^2$ is the metric on a $p$-dimensional compact space $\mathcal{M}_p$. Importantly, we will assume that this compact space may have a non-trivial dependence with respect to the radial coordinate $z$, such that its volume factorizes as
\be
\int d^p\Omega_p=V_{\Omega}(z)\times\text{Vol}(\tilde{\mathcal{M}}_p)\,,
\ee
for a suitable ``normalized'' $\tilde{\mathcal{M}}_p$.

Following the derivation of the previous section, we first consider the scalings
\be
v\equiv \Lambda\,\tau\,, \qquad x_i \equiv \Lambda \, \xi_i \,,\qquad z\equiv \zeta\,,
\ee
while keeping the coordinates of the compact space untouched. Defining $\xi\equiv\xi_i$, $\vec{\xi}_\perp\equiv\xi_j$ $(j\neq i)$ and parameterizing the RT surface with functions $\zeta(\tau,\vec{\xi}_\perp)$,  $\xi(\tau,\vec{\xi}_\perp)$ (assuming no dependence with respect to the coordinates of the compact space), one finds that in the limit $\beta/\Lambda\to 0$ the entanglement entropy functional reduces to:
\bea\label{EntropyAniso}
S_A=\frac{\Lambda^{d-1}\text{Vol}(\tilde{\mathcal{M}}_p)}{4G_N}  \int d\tau d\vec{\xi}_\perp \sqrt{P}\sqrt{Q}\,,
\eea
with
\be
P\equiv V_{\Omega}^2(\zeta)\Pi_{i=1}^{d-1}g_{ii}(\zeta)\,,\qquad Q\equiv (\p_\tau \xi)^2- g_{vv}(\zeta)\left[\frac{1}{g_{ii}(\zeta)}+\sum_{j\neq i}\frac{(\p_{\xi_j} \xi)^2}{g_{jj}(\zeta)}\right]\,.
\ee
In this limit the equation of motion for $\zeta$ is algebraic; however, it generally cannot be written in the form
\be\label{eqform}
v^2=c(\zeta)\,.
\ee
This is due to the anisotropy in the metric (\ref{metric2}). To move forward we specialize to the case of infinite strips, so that the embedding functions are independent of the transverse coordinates, i.e., $\zeta(\tau)$,  $\xi(\tau)$ and hence $\p_{\xi_j}\xi=0$. In this case we recover (\ref{EntropyAniso}) but now with
\be
P\equiv V_{\Omega}^2(\zeta)\Pi_{i=1}^{d-1}g_{ii}(\zeta)\,,\qquad Q\equiv (\p_\tau \xi)^2-\frac{g_{vv}(\zeta)}{g_{ii}(\zeta)}\,.
\ee
The equation of motion for $\zeta$ is now:
\be\label{EQMem}
(\p_\tau \xi)^2=\frac{g_{vv}(\zeta )}{g_{ii}(\zeta )}-\frac{P(\zeta )}{P'(\zeta )g_{ii}(\zeta )^2}\left[g_{vv}(\zeta ) g_{ii}'(\zeta )-g_{ii}(\zeta ) g_{vv}'(\zeta )\right] \,,
\ee
which is of the form (\ref{eqform}) with
\be\label{vcdefs2}
v^2\equiv (\p_\tau \xi)^2\,,\qquad c(\zeta)\equiv  \frac{g_{vv}(\zeta )}{g_{ii}(\zeta )}-\frac{P(\zeta )}{P'(\zeta )g_{ii}(\zeta )^2}\left[g_{vv}(\zeta ) g_{ii}'(\zeta )-g_{ii}(\zeta ) g_{vv}'(\zeta )\right]\,.
\ee
As a consistency check, we note that for asymptotically AdS, isotropic black branes we have
\be
P(\zeta)=\frac{L^{2(d-1)}}{\zeta^{2(d-1)}}\,,\qquad g_{vv}(\zeta)=\frac{L^2}{\zeta^2}a(\zeta)\,,\qquad g_{ii}(\zeta)=\frac{L^2}{\zeta^2}\,,\qquad g_{jj}(\zeta)=\frac{L^2}{\zeta^2}
\ee
and we recover (\ref{vcdefs}) from (\ref{vcdefs2}). Plugging back into the action the solution to this equation, $\zeta=c^{-1}(v^2)$, we can rewrite the entropy functional as a membrane functional:
\bea
S_A=\frac{\Lambda^{d-1}\text{Vol}(\tilde{\mathcal{M}})}{4G_N} \int d\tau d\vec{\xi}_\perp \,{\cal \tilde{E}}\left(v\right)
=\frac{\Lambda^{d-1}\text{Vol}(\tilde{\mathcal{M}})}{4G_N} \int d\text{area}\ { {\cal \tilde{E}}\left(v\right)\over \sqrt{1-v^2}}\,,
\eea
where
\be
d\text{area}=d\tau d\vec{\xi}_\perp\sqrt{1-(\p_\tau \xi)^2}
\ee
and
\be\label{tension}
 {\cal \tilde{E}}\left(v\right)\equiv  {\sqrt{-\frac{P(\zeta )^2}{P'(\zeta )g_{ii}(\zeta )^2}\left[g_{vv}(\zeta ) g_{ii}'(\zeta )-g_{ii}(\zeta ) g_{vv}'(\zeta )\right]}}\Bigg\vert_{\zeta=c^{-1}(v^2)}\,.
\ee
We note that ${\cal \tilde{E}}\left(v\right)$ still needs to be normalized. In order to do so, we identify the entropy density as:
\be
s_{\text{th}}=\frac{\text{Vol}(\tilde{\mathcal{M}})}{4G_N}\sqrt{V_{\Omega}^2(\zeta_h) \Pi_{i=1}^{d-1}g_{ii}(\zeta_h)}=\frac{\text{Vol}(\tilde{\mathcal{M}})}{4G_N}\sqrt{P(\zeta_h)}\,.
\ee
Hence, we find that
\bea
S_A=s_{\text{th}}\Lambda^{d-1} \int d\text{area}\ { {\cal E}\left(v\right)\over \sqrt{1-v^2}}\,,
\eea
with
\be
{\cal E}\left(v\right)\equiv  {\sqrt{-\frac{P(\zeta )^2}{P(\zeta_h )P'(\zeta )g_{ii}(\zeta )^2}\left[g_{vv}(\zeta ) g_{ii}'(\zeta )-g_{ii}(\zeta ) g_{vv}'(\zeta )\right]}}\Bigg\vert_{\zeta=c^{-1}(v^2)}\,.
\ee

\subsubsection*{Information velocity from the membrane theory}

We can include the effects of the entanglement fraction $f$ \cite{Couch:2019zni} in the membrane theory in a manner developed by Mark Mezei, who obtained the results of this section in unpublished work \cite{Mezei:note}. The inclusion of $f$ amounts to the introduction of a penalty factor for the lower end of the membrane:
\be\label{SEEf}
S_A(A(\tau))=s_{\text{th}}\left(\int d\text{area}\ { {\cal E}\left(v\right)\over \sqrt{1-v^2}}+ f \text{vol} (A'(0))\right)\,.
\ee
Thus, the membrane is anchored on $\partial A(\tau)$ and $\partial A'(0)$ on the upper and lower boundaries respectively.  Since the latter is just a boundary term, it does not affect the equations of motion. For the case of strips of width $2R$ one gets membranes with constant $v$, and (\ref{SEEf}) becomes
\be
S_A(A(\tau))=s_{\text{th}} \text{area}(\partial A)(\mathcal{E}(v)\tau+f(R-v\tau))\,
\ee
whose minimum in $v$ is at
\be\label{fveq}
f=\mathcal{E}'(v)\,,
\ee
independently of $\tau$. The entropy saturates when $S_A(A(\tau_{\text{sat}})) =s_{\text{th}} \text{area}(\partial A)R$, which gives
\be\label{v_Ifinal}
v_I=\frac{R}{\tau_{\text{sat}}}=\frac{\mathcal{E}(v)-\mathcal{E}'(v)v}{1-f}\bigg|_{\mathcal{E}'(v)=f}=\frac{\Gamma(f)}{1-f}\,,
\ee
where $\Gamma(f)$ is defined as the Legendre  transform of $\mathcal{E}(v)$. It can be checked that $v_I(f= 0) =v_E$ and $v_I(f= 1) =v_B$.

\section{Information propagation in spin chains with variable nonlocality}
\label{sec: spin chains}

We now turn to the question of how the rate of information spreading in chaotic quantum systems is affected by the presence of nonlocality in various forms.  We begin with the base Hamiltonian $H_L^{(0)}$ of equation \eqref{eq: H local}.  By adding next-nearest-neighbor terms (or next-next-nearest-neighbor terms, etc...) with variable couplings, we introduce a form of mild nonlocality into the Hamiltonian and investigate the affect of these terms on information spreading.\footnote{Spin chains with next-nearest neighbor interactions or related couplings have been studied in many other contexts, see for example \cite{PhysRevB.48.3240,Gu_2004,Liu_2006ent,Kwek_2009,2009PhRvB.79a4439O}.  Entanglement growth in integral models with a form of nonlocality induced by a Lifschitz scaling has been studied in \cite{Mozaffar:2018vmk,Mozaffar:2021nex}. }  More precisely, our total Hamiltonian will be

\begin{equation}\begin{split}\label{eq: H total}
		H_L=H_L^{(0)}+H_L^{(\text{nonlocal})}
\end{split}\end{equation}
where
\begin{equation}\begin{split}\label{eq: H nonlocal}
		H_L^{(\text{nonlocal})}=
		 &J_{zoz}\sum_{i=1}^{L-2}\sigma_z^{(i)}\sigma_z^{(i+2)}
		+J_{zooz}\sum_{i=1}^{L-3}\sigma_z^{(i)}\sigma_z^{(i+3)}+...\\
		 &J_{xox}\sum_{i=1}^{L-2}\sigma_x^{(i)}\sigma_x^{(i+2)}
		+J_{xoox}\sum_{i=1}^{L-3}\sigma_x^{(i)}\sigma_x^{(i+3)}+...~.
\end{split}\end{equation}
All the individual terms in $H_L^{(\text{nonlocal})}$ are quadratic, in the sense that they contain two active sites (i.e. only two Pauli operators per summed term).  The labelling scheme for nonlocal coefficients should be clear: the number of $o$'s in $J_{xo...ox}$ (for example)  indicates the number of spaces between active $\sigma_x$ sites.  In most cases, we will set only a subset of the couplings $J_{xox}, J_{zoz}, J_{xoox}, J_{zooz}...$ to nonzero values simultaneously, however we do allow them to be on the same order as the $J_{zz}=1$ coupling in $H_L^{(0)}$, so they do not generally constitute small perturbations to $H^{(0)}$.  In fact, in our scans below we choose to consider nonlocal couplings up to precisely $J_{\text{nonlocal}}=1$ because in some cases going far beyond this causes the Hamiltonian eigenstate level-spacing to deviate from the Wigner-Dyson distribution indicative of a chaotic regime \cite{PhysRevLett.110.084101}, where we wish to remain.

Equation \eqref{eq: H nonlocal} obviously does not represent the unique set of nonlocal terms which could be considered. We choose to work with $\sigma_z \sigma_z$ terms and $\sigma_x \sigma_x$ terms, because unlike $\sigma_y \sigma_y$, such terms leave the states \eqref{eq: f states} at the center of the energy spectrum, regardless of the values of nonlocal coupling coefficients.  These two types of terms also provide clean examples of nonlocal additions which are, respectively, commuting and non-commuting with the nearest neighbor piece of $H^{(0)}_L$, which we speculate may lead to qualitatively different behavior.  

In the following sections we display some results obtained using the methods outlined in section \ref{subsec: tracking vI in spin chains}, to show the effects of nonlocal terms on the rate of information spreading.  

\subsection{Next-nearest neighbor interactions: $J_{xox}$ and $J_{zoz}$}
\label{subsec: next-nearest neighbor scans}

\begin{figure}[t!]
\centering
\includegraphics[width=3in]{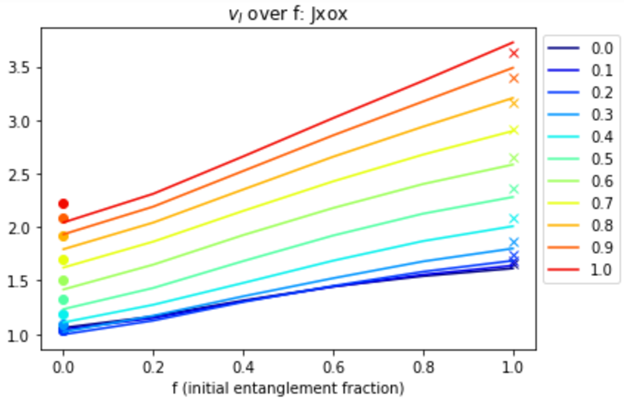}$~$
\includegraphics[width=3in]{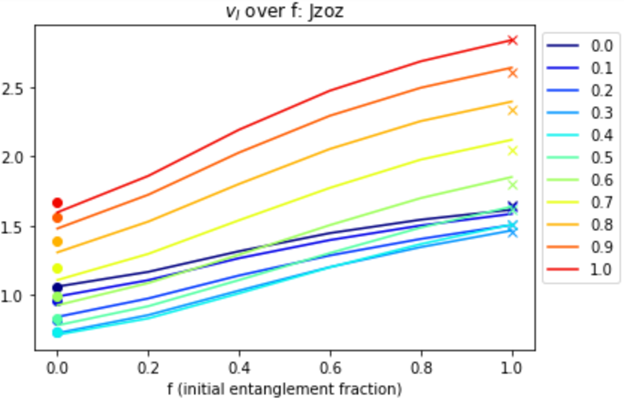}
\begin{picture}(0,0)
\put(-30,145){{\scriptsize $J_{xox}$}}
\put(195,145){{\scriptsize $J_{zoz}$}}
\put(-220,140){{\scriptsize $v_I$}}
\put(5,140){{\scriptsize $v_I$}}
\end{picture}
 \caption{Left: Information velocity as a function of initial entanglement fraction $f$ in $L=26$ systems with different values of $J_{xox}$ nonlocal coupling (all other nonlocal couplings set to zero).  Right: Information velocity as a function of initial entanglement fraction $f$ in $L=26$ systems with different values of $J_{zoz}$ (all other nonlocal couplings set to zero).  In both panels, the $x$'s along the $f=1$ line correspond to independent computations of $v_B$, and the $o$'s along $f=0$ correspond to independent $v_E$ computations. See discussion in the main text. \label{fig: vI over f for Jxox and Jzoz}}
\end{figure}
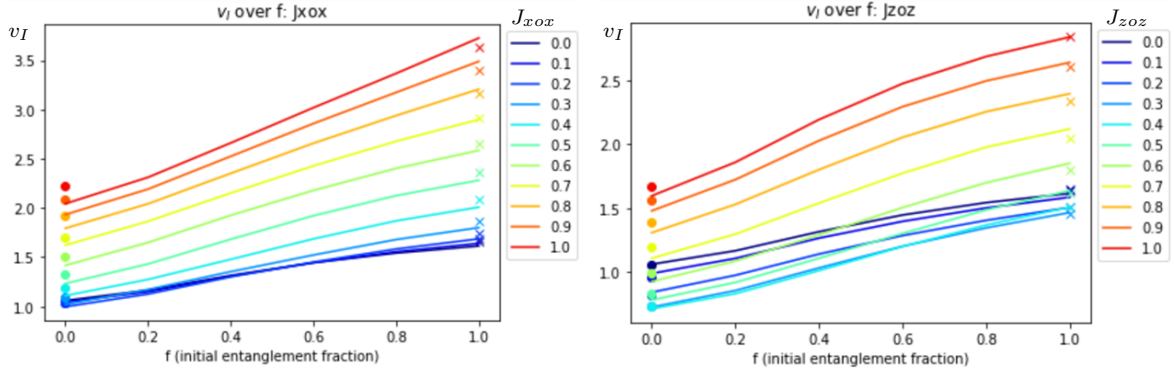

\begin{figure}[t!]
\centering
\includegraphics[width=3in]{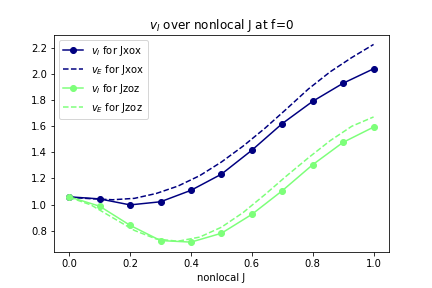}$~$
\includegraphics[width=3in]{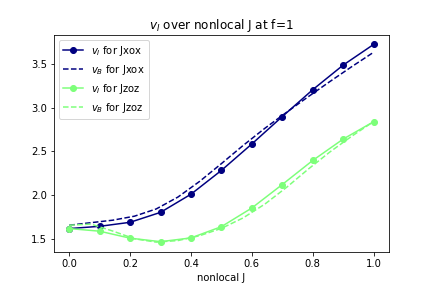}
 \caption{Left: Information velocity as a function of nonlocal coupling ($J_{xox}$ or $J_{zoz}$) at fixed $f=0$. The dashed lines in this panel are an independent computation of $v_E$ values in the presence of the same nonlocal couplings. Right: Information velocity as a function of nonlocal coupling ($J_{xox}$ or $J_{zoz}$) at fixed $f=1$.  The dashed lines in this panel are an independent computation of $v_B$ values in the presence of the same nonlocal couplings. \label{fig: vI over J for Jxox and Jzoz}}
\end{figure}

First we consider the case where only one of the nonlocal terms in \eqref{eq: H nonlocal} is turned on.  Figure \ref{fig: vI over f for Jxox and Jzoz} shows $v_I$ over $f$ for the different values of either $J_{xox}$ (left panel) or $J_{zoz}$ (right panel).  All other nonlocal couplings are set to zero for these scans.  In the case of $J_{xox}$, the most apparent qualitative behavior is perhaps unsurprising: as the amount of nonlocality is increased, the information velocity increases as well, at all values of $f$.  For small $J_{xox}\lesssim 0.2$, however, this effect is suppressed or even reversed, as $v_I$ is either unchanged or even slightly decreased for small $f$. This behavior is far more pronounced in the case of $J_{zoz}$, which shows a clear nonmonotonicity in the behavior of $v_I$ as a function of $J_{zoz}$, at all values of $f$.  This behavior is displayed more clearly in figure \ref{fig: vI over J for Jxox and Jzoz}, and we will discuss it further below.

As a confirmation that the information velocity $v_I(f)$ interpolates between $v_E$ at $f=0$ and $v_B$ at $f=1$, we have computed both of these velocities independently for the same set of system parameters, using basic methods. Results are shown as the dashed lines in figure \ref{fig: vI over J for Jxox and Jzoz}. To compute $v_E$, we tracked the rate of growth entanglement entropy on half of the system, evolving the state $\psi_L^{f=0}$ until saturation (see equation \eqref{eq: f states}) and performing a linear fit on the early linear growth portion of this curve.  The $v_B$ computations consisted of taking the trace of the squared commutator of local Heisenberg operators over a grid of space and time displacements, and finding the rate of growth of the region over which the result is $\mathcal{O}(1)$.  In the case of the $v_E$ computation, the values obtained are particular to the state \eqref{eq: f states} with $f=0$, though they should not substantially differ from those obtained by averaging over random unentangled states.  On the other hand the $v_B$ computation is a state-independent estimation of the butterfly velocity at infinite temperature, as is appropriate for comparisons with states of energy $\expval{H}=0$.  The approximate match of these curves with the behavior of $v_I$, shown in figure \ref{fig: vI over J for Jxox and Jzoz}, is good support for the analysis of \cite{Couch:2019zni} and a good overall consistency check. It is also a confirmation of the non-monotonic behavior seen in the $v_I$ curves, which we now discuss.

\begin{figure}[t!]
\centering
\includegraphics[width=3in]{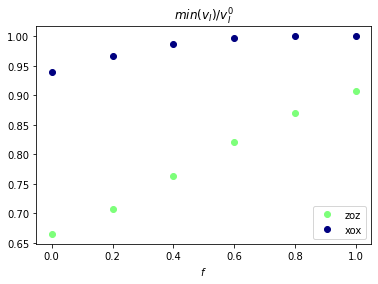}$~$
\includegraphics[width=3in]{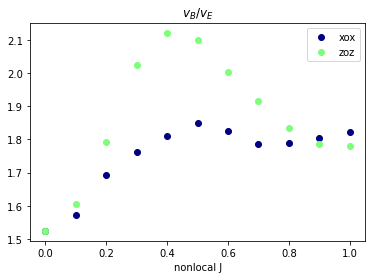}
 \caption{Left: Ratio of the minimum information velocity to the information velocity at zero nonlocal coupling ($J_{xox}=J_{zoz}=0$), denoted $v_I^0$, as a function of entanglement fraction $f$.  Right: Ratio of the butterfly velocity to the entanglement velocity as a function of the nonlocal coupling ($J_{xox}$ or $J_{zoz}$).  \label{fig: velocity ratios}}
\end{figure}

The nonmonotonicity evident in figure \ref{fig: vI over J for Jxox and Jzoz} is somewhat counter-intuitive, as it indicates that in some cases the inclusion of nonlocal terms in the Hamiltonian can suppress the rate of information propagation.  Such behavior is not observed in any of the holographic systems investigated in section \ref{sec: holography}.  For the case of $J_{xox}$ couplings, the suppressive effect is almost small enough to be dismissed as an artifact of finite size effects or procedural uncertainties, with $v_I$ never falling below about 6 percent less than its value at $J_{xox}=0$ (though even ``flatness'' of the $v_I(J_{xox})$ curve for small $J_{xox}$ would be noteworthy).  The case of $J_{zoz}$ shows a much larger suppressive effect, peaking around $J_{zoz}\sim 0.4$, where $v_I$ is nearly 35 percent lower than its value with no nonlocal coupling.  We focus on this effect in the left panel of figure \ref{fig: velocity ratios}, plotting the ratio of the minimum $v_I$ (with respect to $J_{\text{nonlocal}}$) and $v_I$ when $J_{\text{nonlocal}}=0$. The effect is most stark at $f=0$.  

In the right hand side of figure \ref{fig: velocity ratios}, we have plotted the ratio $v_B/v_E$, as inferred from $v_I(f=0)/v_I(f=1)$, at various values of nonlocal coupling.  The precise form of the curves emerging from this data are surely dependent on our specific choices of Hamiltonian and couplings, but we include this plot for comparison with holographic results (see figures \ref{plots1} and \ref{plots1b}, left panels), where the analogous ratios are monotonically increasing and relatively featureless.

\subsection{More nonlocality: fixed width scans}
\label{subsec: fixed width scans}

\begin{figure}[t!]
\centering
\includegraphics[width=3in]{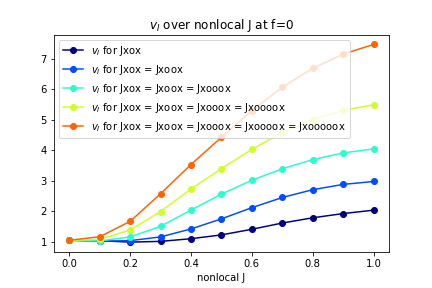}$~$
\includegraphics[width=3in]{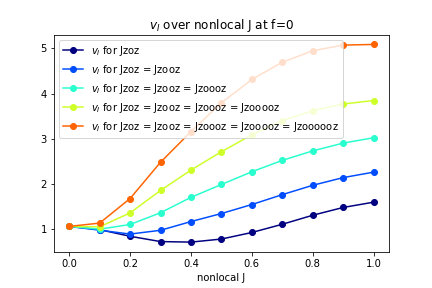}
 \caption{Left: Information velocity as a function of nonlocal $\sigma_x$ couplings, at $f=0$.  Nonlocal terms are turned on up to fixed ``width''. Right: Information velocity as a function of nonlocal $\sigma_z$ couplings, at $f=0$. Nonlocal terms are turned on up to fixed ``width''.  \label{fig: fixed width scans}}
\end{figure}

We now move beyond the case of next-nearest neighbor interactions, and consider adding a higher degree of nonlocality.  We fix all $J_{xox}, J_{xoox}, ..., J_{xo...ox}$ couplings, up to a fixed ``width'', to the same $J_{\text{nonlocal}}$ value and turn them on simultaneously.  Results for $v_I$ as a function of $J_{\text{nonlocal}}$ are displayed in \ref{fig: fixed width scans}.  In this case, the affect of additional nonlocal terms is simply to raise the $v_I(J_{\text{nonlocal}})$ curves, moreso with the inclusion of each additional ``width'' of nonlocal terms.   This also has the affect of washing out the nonmonotonic behavior observed in section \ref{subsec: next-nearest neighbor scans} and approaching behavior more similar to the holographic cases we consider in section \ref{sec: holography}.
It is interesting that the growth of $v_I$ begins to level off near $J_{\text{nonlocal}}=1$.  In principle these coupling parameters could be made arbitrarily large, but we have chosen to limit $J_{\text{nonlocal}} \le 1$ in order to leave a comfortable margin before the Hamiltonian level-spacing departs from the Wigner-Dyson distribution expected of chaotic systems. 

\subsection{Membrane tension function and nonlocality}
\label{subsec: spin chain approx membranes}

\begin{figure}[t!]
\centering
\includegraphics[width=3in]{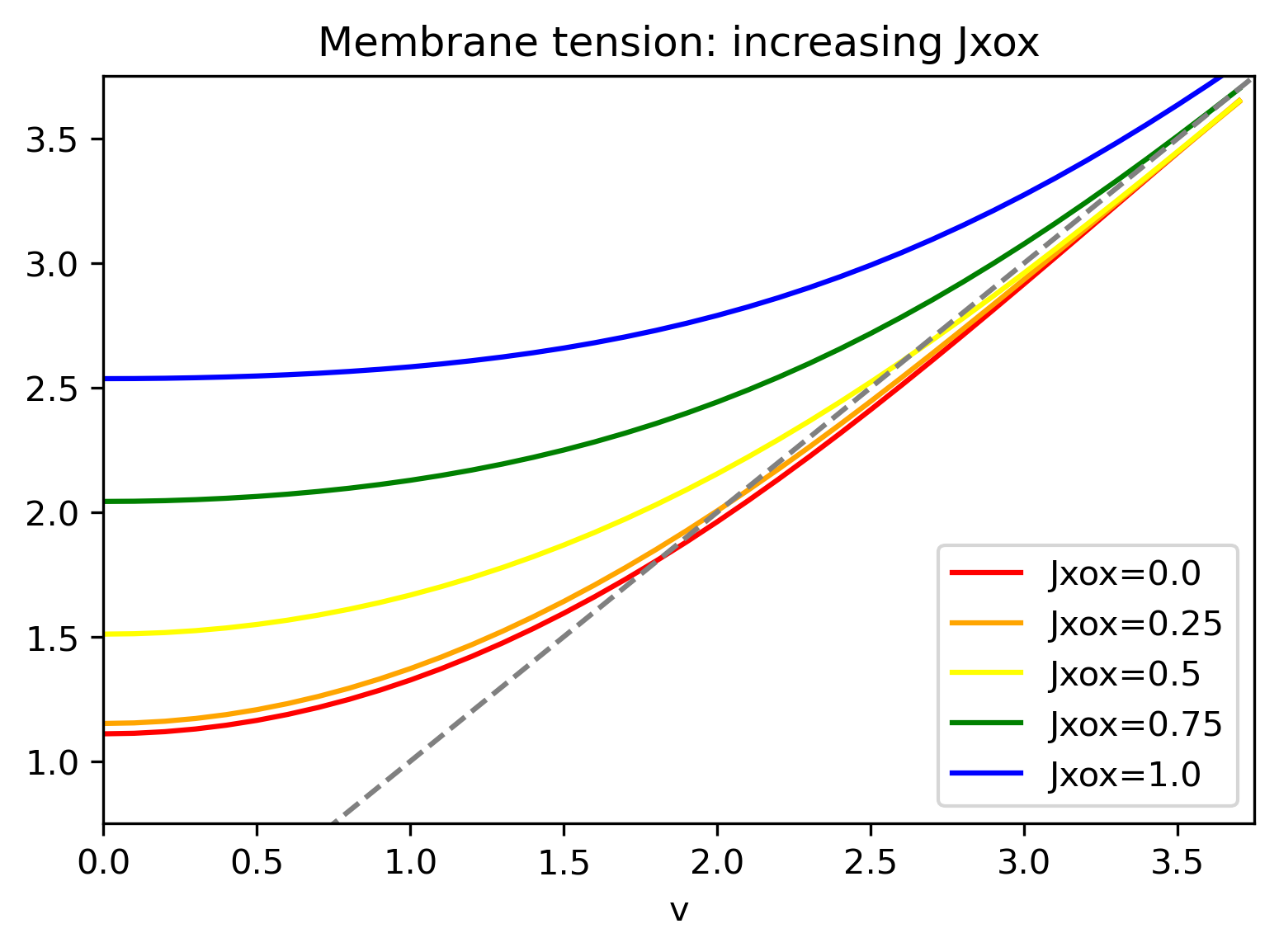}$\qquad$\includegraphics[width=3in]{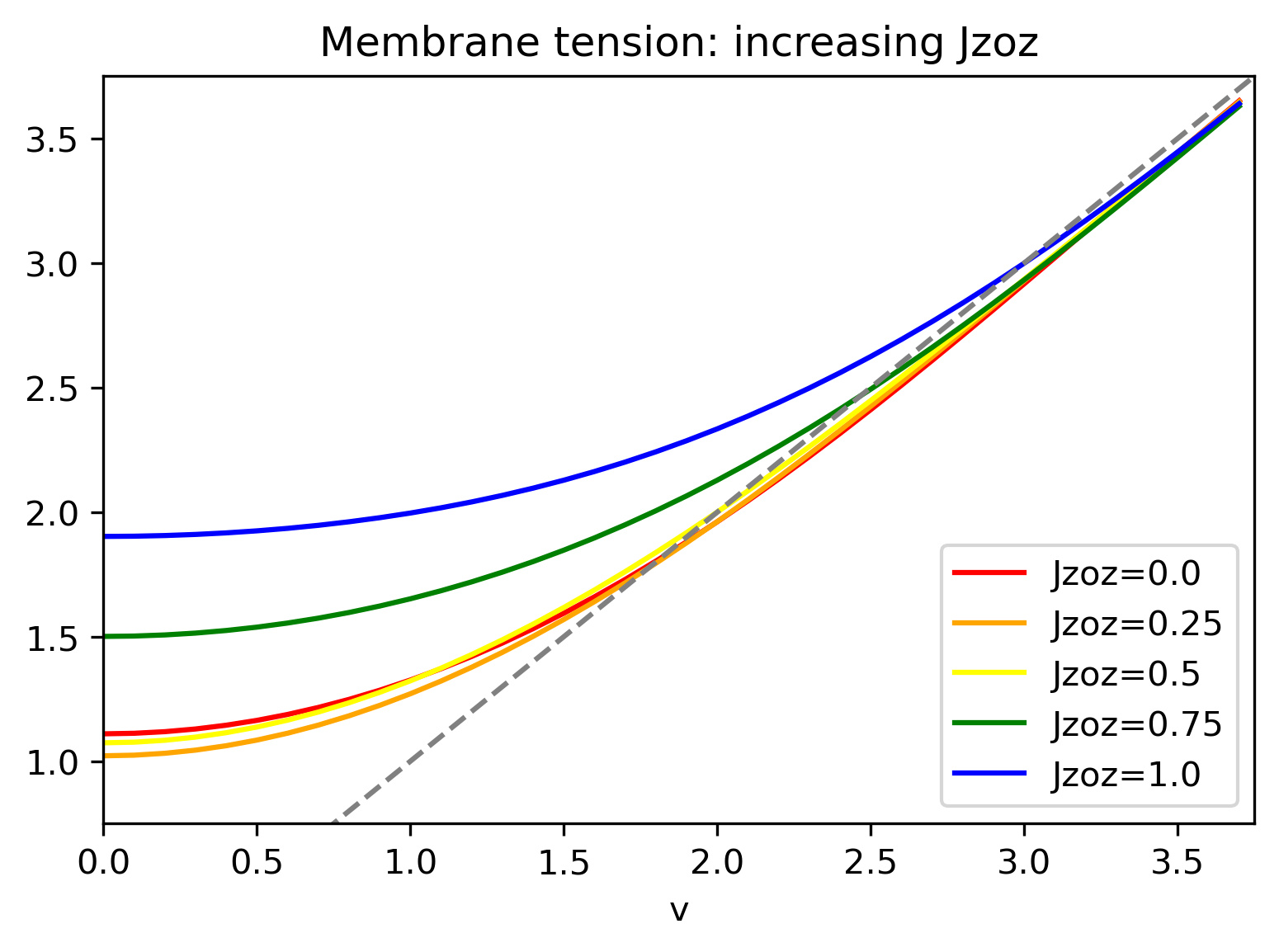}
\begin{picture}(0,0)
\put(-220,160){{\scriptsize ${\cal E}$(v)}}
\put(18,160){{\scriptsize ${\cal E}$(v)}}
\end{picture}
\caption{Numerical estimation of the membrane tension function ${\cal E}(v)$ for various the spin chain with a single nonlocal coupling turned on.  Left: The overall affect of increasing the $J_{xox}$ coupling is to raise the membrane tension function. Right: Increasing the $J_{zoz}$ coupling initially has a small suppressive effect on the membrane tension function, but beyond $J_{zoz}\simeq 0.5$ it surpasses the $J_{zoz}=0$ values.\label{fig: spin chain membranes}}
\end{figure}
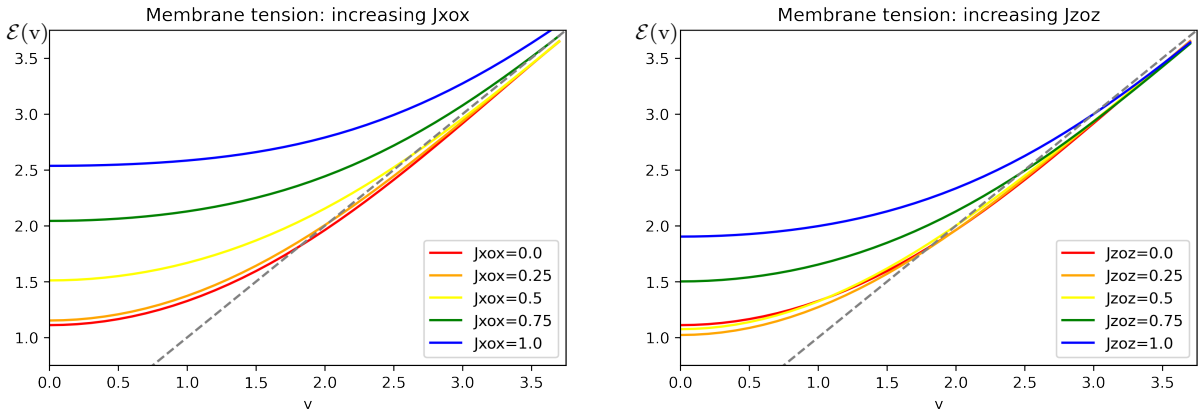

We now approximate the membrane tension function for the systems of section \ref{subsec: next-nearest neighbor scans} with variable next-nearest neighbor couplings, using the method outlined in section \ref{subsubsec: membrane theory in spin chains} with $L=13$. Results are displayed in figure \ref{fig: spin chain membranes}.   Finite size affects are substantial, so these curves can only be considered schematic representations of the effect of the nonlocal coupling on ${\cal E}(v)$ (i.e. they do not give precise values of $v_E={\cal E}(0)$ or $v_B={\cal E}(v_B)$).  They are obtained by computing ${\cal E}_{\text{eff}}(t,v)$ over a series of $t$ values up to the $v_B$ crossing time for each Hamiltonian and fitting the boundary of the resulting data set with an even polynomial of degree 12.  The central portion (toward $v=0$) of the curves, which in an infinite system would converge through late time values of ${\cal E}_{\text{eff}}(t,v)$, tend to be overestimated.  This effect is larger for higher (true) values of $v_E, v_B$, because finite size effects prohibit a good estimate from ${\cal E}_{\text{eff}}(t,v)$ at much lower values of $t$.
All these caveats aside, in qualitative terms we find good agreement with the results of the previous section: the effect of adding $J_{xox}$ is initially fairly flat, but past $J_{xox}\sim 0.25$ dramatically raises ${\cal E}(v)$ upward at all $v$ (entailing an increase of $v_B$, $v_E$, and $v_I$).  Increasing $J_{zoz}$ from zero initially has a small suppressive effect, but by $J_{zoz}\sim 0.5$ the curve raises to match original values ($J_{zoz}=0$ values) and continues to increase uniformly.  

\section{Information propagation in nonlocal holographic systems}
\label{sec: holography}

Inspired by the lattice results of the previous section, we now turn our attention to large-$N$ gauge theories with holographic duals. We will consider two prototypical theories with different types of nonlocality: non-commutative SYM (NCSYM) theory and dipole-deformed SYM (DDSYM) theory. The former is an example of a nonlocal theory with UV/IR mixing. A simple model that realizes non-commutativity consists of a set of oppositely charged particles (dipoles) moving in a strong magnetic field \cite{Bigatti:1999iz}. The UV/IR mixing then implies that the transverse size of the dipoles grows with their longitudinal momentum, which may in turn posit obstructions for renormalizability \cite{Minwalla:1999px,VanRaamsdonk:2000rr}. Even though this theory is ultimately finite, noncommutativity is arguably not the simplest way to introduce nonlocality. We thus consider a second theory, inspired by \cite{Bergman:2000cw}, which introduces a set of fundamental dipoles of \emph{constant} length $L$, hence, resembling more closely the lattice systems studied in section \ref{sec: spin chains}. As a result, one obtains a nonlocal theory without the issue of UV/IR mixing. 

In the following, we will review the basic features of the holographic duals of NCSYM and DDSYM, realized as top-down constructions in string theory. We will also comment on their finite temperature black brane solutions, which are needed to tackle the questions pertaining to thermalization and information spreading.\footnote{To our understanding, this is the first time that a black brane solution is derived for the gravity dual of DDSYM theory.} Finally, we will use the tools developed in \ref{subsec: membrane theory in holography} to write down membrane tension functions for both systems and analyze the results.

\subsection{Theories of interest}

\subsubsection{Gravity dual of NCSYM}

The gravity dual of maximally supersymmetric non-commutative SYM (NCSYM) is given by type IIB supergravity with non-zero NS-NS $B$-field. Here and below we assume that the non-commutative parameter is non-vanishing only in the $(x^2,x^3)$-plane, i.e., $[x^2,x^3]\sim i \theta$. This amounts to replace all multiplication in the Lagrangian of the $\mathcal{N}=4$ SYM theory with a noncommutative star product
\be
(f\star g)(x_2,x_3)=e^{\frac{i}{2}\theta\left(\partial_{\xi_2}\partial_{\zeta_3}-\partial_{\xi_3}\partial_{\zeta_2}\right)}f(x_2+\xi_2,x_3+\xi_3)g(x_2+\zeta_2,x_3+\zeta_3)|_{\xi_2=\xi_3=\zeta_2=\zeta_3=0}\,.
\ee
The gravity background dual to this theory, in the string frame is \cite{Hashimoto:1999ut,Maldacena:1999mh}:
\begin{eqnarray}\label{backg}
ds_\mt{st}^2 &=&R^2 \left[r^2\left(-f(r) dt^2+dx_1^2+h(r)\left(dx_2^2+dx_3^2\right)\right)+ \frac{dr^2}{r^2f(r)}+ d\tilde{\Omega}_5^2\right],\nonumber\\
e^{2\Phi} &=& {\hat g}^2h(r)\,,\nonumber\\
B_{23} &=& R^2a^2 r^4 h(r)\,,\\
C_{01} &=& \frac{R^2a^2}{\hat g}r^4\,,\nonumber\\
F_{0123r} &=& \frac{4R^4}{{\hat g}}r^3 h(r)\,,\nonumber
\end{eqnarray}
with
\be
f(r)=1-\left(\frac{r_h}{r}\right)^4\,,\qquad r_h=\pi T\,,
\ee
and
\be
h(r)=\frac{1}{1+a^4r^4}\,,\qquad a=\lambda^{1/4}\sqrt{\theta}\,.
\ee
In the above $d\tilde{\Omega}_5^2$ denotes the metric on a unit $S^5$, $R^4 = 4 \pi \hat g N\alpha'^2$, $\hat g$ denotes the string coupling and $\alpha'$ is the string tension, which is related to the 't Hooft coupling through the standard relation $\sqrt{\lambda}=R^2/\alpha'$.

To use the results for the membrane theory we first express the metric in Einstein frame,\footnote{The metrics in the Einstein and string frames are related via the standard relation: $ds_\mt{E}^2=e^{-\Phi/2}ds_\mt{st}^2$.}
\be
ds_\mt{E}^2 =\frac{ R^2}{{\hat g}^{1/2}h(r)^{1/4}} \left[r^2\left(-f(r) dt^2+dx_1^2+h(r)\left(dx_2^2+dx_3^2\right)\right)+ \frac{dr^2}{r^2f(r)}+ d\tilde{\Omega}_5^2\right]\,,
\ee
and go to a radial coordinate $z=1/r$, so that:
\be
ds_\mt{E}^2=\frac{ R^2}{{\hat g}^{1/2}h(z)^{1/4}z^2} \left[-f(z) dt^2+dx_1^2+h(z)\left(dx_2^2+dx_3^2\right)+ \frac{dz^2}{f(z)}+ z^2d\tilde{\Omega}_5^2\right]\,,
\ee
where
\be
f(z)=1-\left(\frac{z}{z_h}\right)^4\,,\qquad z_h=\frac{1}{\pi T}\,,
\ee
and
\be
h(z)=\frac{1}{1+\left(\frac{a}{z}\right)^4}\,,\qquad a=\lambda^{1/4}\sqrt{\theta}\,.
\ee
We further define the infalling time coordinate $v$ such that
\be
dt=dv+\frac{dz}{f(z)}\,,
\ee
hence the metric becomes:
\be
ds_\mt{E}^2=\frac{ R^2}{{\hat g}^{1/2}h(z)^{1/4}z^2} \left[-f(z) dv^2-2dvdz+dx_1^2+h(z)\left(dx_2^2+dx_3^2\right)+ z^2d\tilde{\Omega}_5^2\right]\,.
\ee
This metric is of the form (\ref{metric2}), with the following identifications:
\be
g_{vv}=\frac{ R^2f(z)}{{\hat g}^{1/2}h(z)^{1/4}z^2}\,,\quad g_{vz}=g_{11}=\frac{ R^2}{{\hat g}^{1/2}h(z)^{1/4}z^2}\,,\quad g_{22}=g_{33}=\frac{ R^2h(z)^{3/4}}{{\hat g}^{1/2}z^2}\,.
\ee
Furthermore, the metric of the compact space is
\be
d\Omega_5^2=\frac{ R^2}{{\hat g}^{1/2}h(z)^{1/4}}d\tilde{\Omega}_5^2\,,
\ee
so we obtain that
\be
V_{\Omega}(z)=\frac{R^5}{{\hat g}^{5/4}h(z)^{5/8}}\,,\qquad \text{Vol}(\tilde{\Omega}_5)=\pi^3\,.
\ee
With the above identifications, we are now ready to apply our general results of section \ref{subsec: membrane theory in holography} specializing to this system.

\subsubsection{Gravity dual of DDSYM}

The gravity dual of the dipole deformed maximally supersymmetric SYM (DDSYM) is given by type IIB supergravity with non-zero NS-NS $B$-field and a deformed compact space. Here and below we assume that the dipoles have fixed length $L$ and are all oriented along the $x^3$-direction. This amounts to replace all multiplication in the Lagrangian of the $\mathcal{N}=4$ SYM theory with the star product
\be
(f\star g)(x_3)=f\left(x_3-\frac{L}{2}\right)g\left(x_3+\frac{L}{2}\right)\,.
\ee
The gravity background dual to this theory, in the string frame is \cite{Bergman:2001rw}:\footnote{The metric presented in \cite{Bergman:2001rw} is the zero temperature version of (\ref{backg}), where $f(r)=1$. To our understanding, this is the first time that a black brane solution was derived for the gravity dual of the dipole theory.}
\begin{eqnarray}\label{backg}
ds_\mt{st}^2 &=&R^2 \left[r^2\left(-f(r) dt^2+dx_1^2+dx_2^2+h(r)dx_3^2\right)+ \frac{dr^2}{r^2f(r)}+ d\mathrm{\bar{\Omega}}_5^2\right],\nonumber\\
e^{2\Phi} &=& {\hat g}^2h(r)\,,\nonumber\\
B_{3\psi} &=& R^2b\, r^2 h(r)\,,\\
F_{0123r} &=& 4R^4r^3 \,,\nonumber
\end{eqnarray}
with
\be
f(r)=1-\left(\frac{r_h}{r}\right)^4\,,\qquad r_h=\pi T\,,
\ee
and
\be\label{hrdipole}
h(r)=\frac{1}{1+b^2r^2}\,,\qquad b=\frac{\lambda^{1/2}L}{2\pi}\,.
\ee
In the above $d\bar{\Omega}_5^2$ denotes the metric on a \emph{deformed} unit $S^5$. This compact space has the structure of an $S^1$ (Hopf) fibration over a base $\mathbb{C}\mathbb{P}^2$. The global angular 1-form of the Hopf fibration is denoted as $\psi$. This fiber acquires an $r$-dependent radius $h(r)^{1/2}$.
The volume of the $\mathbb{C}\mathbb{P}^2$ is constant and given by $\pi^2/2$, so the total volume of the compact manifold is $\pi^3h(r)^{1/2}$.

To use the results for the membrane theory we first express the metric in the Einstein frame
\be
ds_\mt{E}^2 =\frac{ R^2}{{\hat g}^{1/2}h(r)^{1/4}} \left[r^2\left(-f(r) dt^2+dx_1^2+dx_2^2+h(r)dx_3^2\right)+ \frac{dr^2}{r^2f(r)}+ d\mathrm{\bar{\Omega}}_5^2\right]\,,
\ee
and go to a radial coordinate $z=1/r$:
\be
ds_\mt{E}^2=\frac{ R^2}{{\hat g}^{1/2}h(z)^{1/4}z^2} \left[-f(z) dt^2+dx_1^2+dx_2^2+h(z)dx_3^2+ \frac{dz^2}{f(z)}+ z^2d\bar{\Omega}_5^2\right]\,,
\ee
where
\be
f(z)=1-\left(\frac{z}{z_h}\right)^4\,,\qquad z_h=\frac{1}{\pi T}\,,
\ee
and
\be
h(z)=\frac{1}{1+\left(\frac{b}{z}\right)^2}\,,\qquad b=\frac{\lambda^{1/2}L}{2\pi}\,.
\ee
We further define the infalling time coordinate $v$ such that
\be
dt=dv+\frac{dz}{f(z)}\,,
\ee
hence the metric becomes:
\be
ds_\mt{E}^2=\frac{ R^2}{{\hat g}^{1/2}h(z)^{1/4}z^2} \left[-f(z) dv^2-2dvdz+dx_1^2+dx_2^2+h(z)dx_3^2+ z^2d\bar{\Omega}_5^2\right]\,.
\ee
This metric is of the form (\ref{metric2}), with the following identifications:
\be
g_{vv}=\frac{ R^2f(z)}{{\hat g}^{1/2}h(z)^{1/4}z^2}\,,\quad g_{vz}=g_{11}=g_{22}=\frac{ R^2}{{\hat g}^{1/2}h(z)^{1/4}z^2}\,,\quad g_{33}=\frac{ R^2h(z)^{3/4}}{{\hat g}^{1/2}z^2}\,.
\ee
Furthermore, the metric of the compact space is
\be
d\Omega_5^2=\frac{ R^2}{{\hat g}^{1/2}h(z)^{1/4}}d\bar{\Omega}_5^2\,.
\ee
From the discussion below (\ref{hrdipole}) we know that $\text{Vol}(\bar{\Omega}_5)=\pi^3h(z)^{1/2}$. Extracting all $z$-dependent terms into $V_{\Omega}(z)$,
we obtain that
\be
V_{\Omega}(z)=\frac{R^5}{{\hat g}^{5/4}h(z)^{1/8}}\,,\qquad \text{Vol}(\tilde{\Omega}_5)=\pi^3\,.
\ee
Again, with these identifications, we can now apply our general results of section \ref{subsec: membrane theory in holography} specializing to this system.

\subsection{Membrane theory and results for information spreading}

\subsubsection{Membrane theory for NCYSM}

We can analyze two cases, depending on the orientation of the strip. In both cases we find:
\be
P(\zeta)=\frac{R^{16}}{{\hat g}^{4}\zeta^6}\,,
\ee
however, the membrane tensions differ.
\begin{itemize}
  \item \textbf{Commutative strip} ($\xi=\xi_1$ and $\vec{\xi}_\perp=\{\xi_2,\xi_3\}$)
\end{itemize}
First, note that (\ref{EQMem}) can be written as:
\be
v^2=c_{\text{c}}(\zeta)\,,\qquad c_{\text{c}}(\zeta)=f(\zeta)-\frac{1}{6} \zeta f'(\zeta)=1-\frac{\zeta^4}{3 \zeta_h^4}\,,
\ee
and this equation can be inverted analytically to give:
\be
\zeta_\text{c}=c_{\text{c}}^{-1}(v^2)=\zeta_h\left[3 (1- v^2)\right]^{1/4}\,.
\ee
We have added a subindices ``c'' to indicate that we are working out the commutative strip. Finally, the membrane tension is
\be\label{membraneC}
{\cal E}_{\text{c}}\left(v\right)=\zeta_h^3 \sqrt{-\frac{f'(\zeta)}{6 \zeta^5}}\Bigg\vert_{\zeta_\text{c}}=\sqrt{\frac{2}{3}}\frac{\zeta_h}{\zeta}\Bigg\vert_{\zeta_\text{c}}=\frac{\sqrt{2}}{3^{3/4} (1-v^2)^{1/4}}\,.
\ee
This is exactly the same result for a strip in Schwarzschild-AdS$_{5}$. It satisfies all the desired properties, including:
\be
 {\cal E}_{\text{c}}(0)=v_E \,, \qquad {\cal E}_{\text{c}}'(0)=0\,, \qquad {\cal E}_{\text{c}}(v_B)=v_B\,, \qquad {\cal E}_{\text{c}}'(v_B)=1\,,
\ee
with
\be
v_E^{\text{(c)}}=\frac{\sqrt{2}}{3^{3/4}}\,,\qquad v_B^{\text{(c)}}=\sqrt{\frac{2}{3}}\,.
\ee

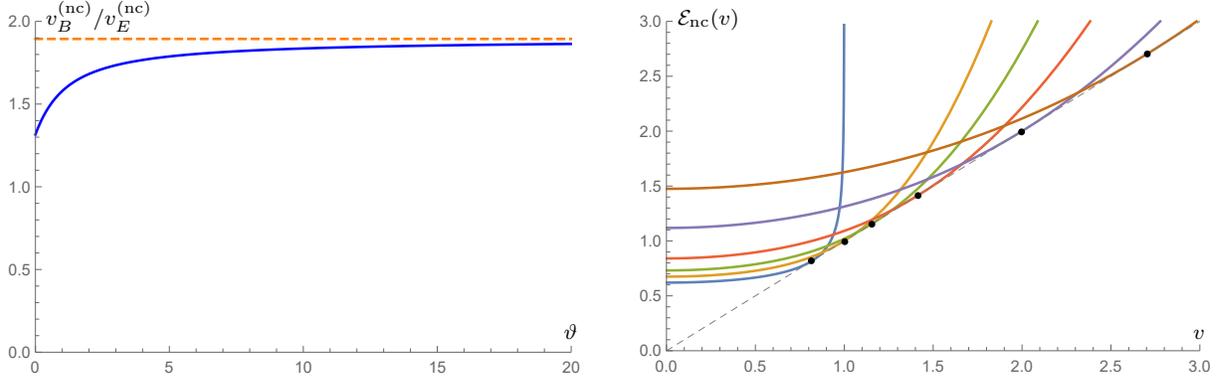
\begin{figure}[t!]
\centering
\includegraphics[width=3in]{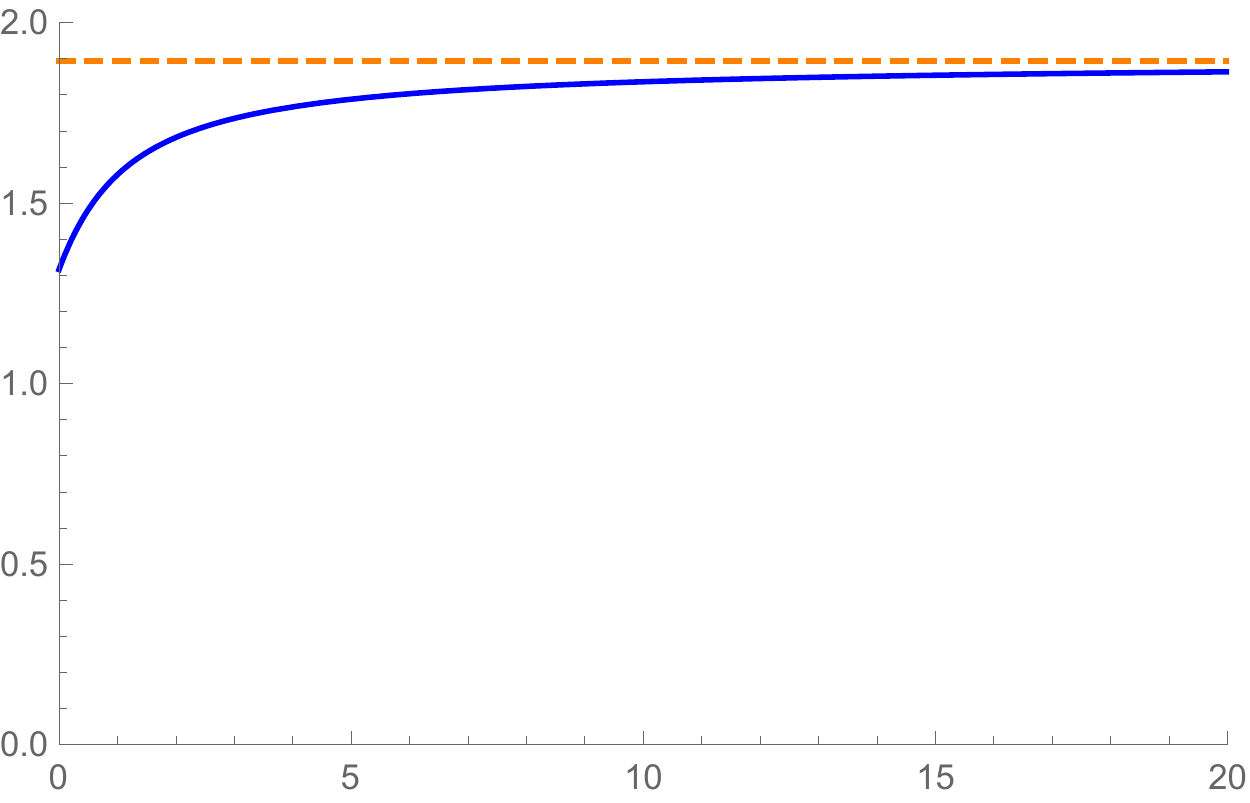}$\qquad$\includegraphics[width=3in]{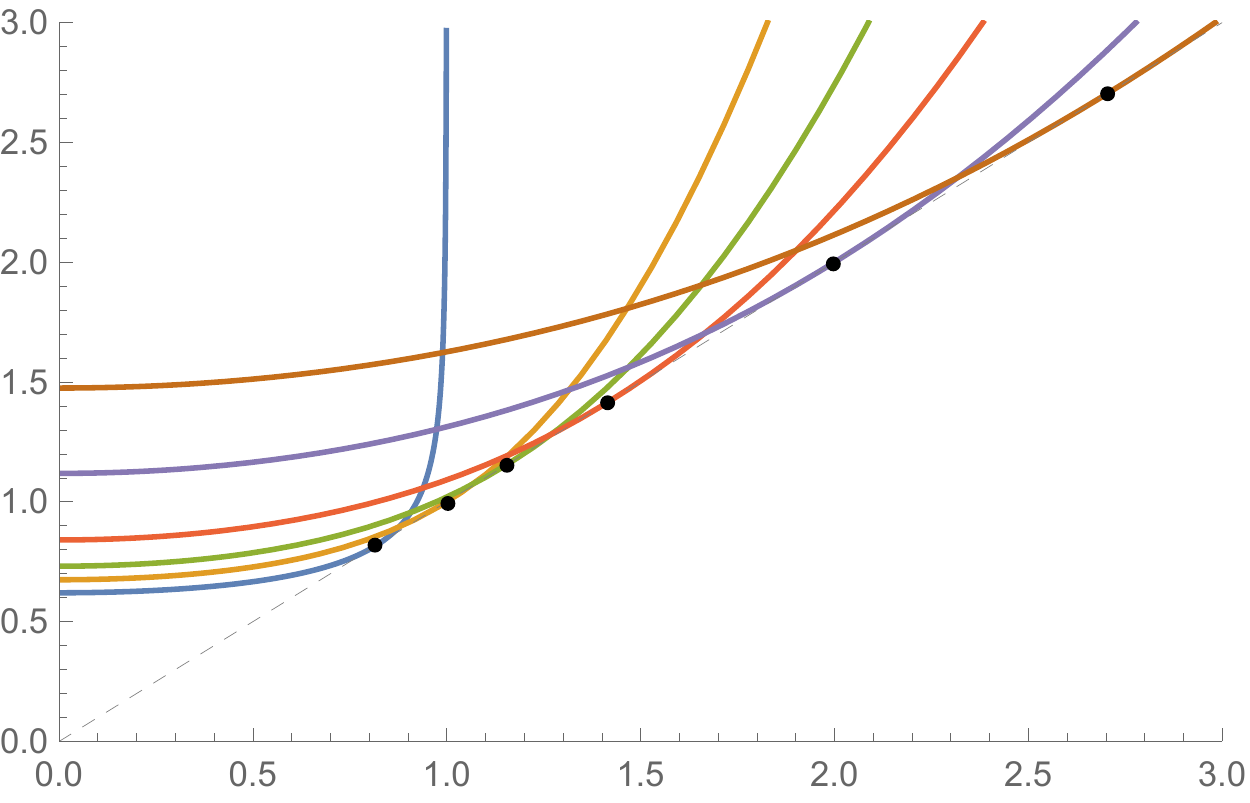}
  \begin{picture}(0,0)
\put(-207,149){{\scriptsize $v_B^{\text{(nc)}}/v_E^{\text{(nc)}}$}}
\put(32,149){{\scriptsize $\mathcal{E}_{\text{nc}}(v)$}}
\put(-11,30){{\scriptsize $\vartheta$}}
\put(227,30){{\scriptsize $v$}}
\end{picture}
 \caption{Left: ratio between the butterfly velocity and the entanglement velocity as a function of the dimensionless combination $\vartheta=\lambda\pi^4T^4\theta^2$. We also show the asymptotic value $v_B/v_E\to (5/3)^{5/4}\approx1.894$, depicted in orange. Right: Membrane tensions for $\vartheta\in\{0,1/2,1,2,5,10\}$, from bottom to top, respectively. In each case, we indicate the point at which the curve touches the straight line $\mathcal{E}(v_B)=v_B$.\label{plots1}}
\end{figure}

\begin{itemize}
  \item \textbf{Non-commutative strip} ($\xi=\xi_2$ and $\vec{\xi}_\perp=\{\xi_1,\xi_3\}$)
\end{itemize}
In this case (\ref{EQMem}) can be written as:
\be
v^2=c_{\text{nc}}(\zeta)\,,\qquad c_{\text{nc}}(\zeta)=\frac{f(\zeta)}{h(\zeta)}-\frac{\zeta f'(\zeta)}{6 h(\zeta)}+\frac{\zeta f(\zeta) h'(\zeta)}{6 h(\zeta)^2}=1-\frac{\zeta^4}{3 \zeta_h^4}+\frac{5 a^4}{3 \zeta^4}-\frac{a^4}{\zeta_h^4}\,,
\ee
and can also be inverted analytically to yield:\footnote{There are 8 roots (4 imaginary and 4 real), but only one coincides with the commutative case in the limit $a\to0$.}
\be
\zeta_\text{nc}=c_{\text{nc}}^{-1}(v^2)=\frac{\zeta_h}{2^{1/4}}\left[3\left(1-v^2-\frac{a^4}{\zeta_h^4}\right)+\sqrt{9\left(1-v^2-\frac{a^4}{\zeta_h^4}\right)^2+\frac{20a^4}{\zeta_h^4}}\right]^{1/4}\,.
\ee
We have added a subindices ``nc'' to indicate that we are working out the non-commutative strip. Finally, the membrane tension is
\bea
{\cal E}_{\text{nc}}\left(v\right)&=&\zeta_h^3 \sqrt{-\frac{f'(\zeta)h(\zeta)-f(\zeta) h'(\zeta)}{6 \zeta^{5} h(\zeta)^2}}\Bigg\vert_{\zeta_{\text{nc}}}=\sqrt{\frac{2}{3} \left(1+\frac{a^4 \zeta_h^4}{\zeta^8}\right)}\frac{\zeta_h}{\zeta}\Bigg\vert_{\zeta_{\text{nc}}}\nonumber\\
&=&\frac{2^{3/4}\!\!\left[\left(3 \left(1-v^2-\frac{a^4}{\zeta_h^4}\right)+\sqrt{9 \left(1-v^2-\frac{a^4}{\zeta_h^4}\right)^2+\frac{20 a^4}{\zeta_h^4}}\right)^2+\frac{4 a^4}{\zeta_h^4}\right]^{1/2}}{\sqrt{3}\left[3 \left(1-v^2-\frac{a^4}{\zeta_h^4}\right)+\sqrt{9 \left(1-v^2-\frac{a^4}{\zeta_h^4}\right)^2+\frac{20 a^4}{\zeta_h^4}}\right]^{5/4}}\,.
\eea
We can check that in the $a\to0$ limit we recover (\ref{membraneC}). It also satisfies all the desired properties:
\be
 {\cal E}_{\text{nc}}(0)=v_E \,, \qquad {\cal E}_{\text{nc}}'(0)=0\,, \qquad {\cal E}_{\text{nc}}(v_B)=v_B\,, \qquad {\cal E}_{\text{nc}}'(v_B)=1\,,
\ee
now with
\be
v_E^{\text{(nc)}}=\frac{2^{3/4}\!\!\left[\left(3 \left(1-\frac{a^4}{\zeta_h^4}\right)+\sqrt{9 \left(1-\frac{a^4}{\zeta_h^4}\right)^2+\frac{20 a^4}{\zeta_h^4}}\right)^2+\frac{4 a^4}{\zeta_h^4}\right]^{1/2}}{\sqrt{3}\left[3 \left(1-\frac{a^4}{\zeta_h^4}\right)+\sqrt{9 \left(1-\frac{a^4}{\zeta_h^4}\right)^2+\frac{20 a^4}{\zeta_h^4}}\right]^{5/4}}\,,\qquad v_B^{\text{(nc)}}=\sqrt{\frac{2}{3}\left(1+\frac{a^4}{\zeta_h^4}\right)}\,.
\ee
Note that in all the above formulas, the non-commutative parameter $\theta$ only appears through the dimensionless combination:
\be
\vartheta\equiv\frac{a^4}{\zeta_h^4}=\lambda\pi^4T^4\theta^2\,.
\ee
We can verify that $v_B^{\text{(nc)}}>v_E^{\text{(nc)}}$ for all $\vartheta$ \cite{Fischler:2018kwt}. Also, both $v_E^{\text{(nc)}}\to\infty$ and $v_B^{\text{(nc)}}\to\infty$ as $\vartheta\to\infty$, but this is the expected behavior for an ``infinitely'' nonlocal theory. Even so, they diverge with the same power of $\vartheta$ ($v_B^{\text{(nc)}}\propto \vartheta^2$, $v_E^{\text{(nc)}}\propto \vartheta^2$) so their ratio remains finite. Expanding this ratio in the limits of strong and weak non-commutativity, we obtain:
\be
\frac{v_B^{\text{(nc)}}}{v_E^{\text{(nc)}}}=\begin{cases}
  \displaystyle 3^{1/4}\left[1+\frac{\vartheta}{3}+\mathcal{O}(\vartheta^2)\right]& \displaystyle \qquad (\vartheta\ll1) \,,\\[3ex]
  \displaystyle \left(\frac{5}{3}\right)^{5/4}\left[1-\frac{1}{3\vartheta}+\mathcal{O}\left(\frac{1}{\vartheta^2}\right)\right] & \displaystyle \qquad (\vartheta\gg1)\,.
 \end{cases}
\ee
These results are illustrated in figure \ref{plots1}. From the plots we can confirm the monotonicity of $v_B^{\text{(nc)}}/v_E^{\text{(nc)}}$ with the strength of the non-commutative parameter. More specifically, the ratio interpolates between $v_B^{\text{(nc)}}/v_E^{\text{(nc)}}\to3^{1/4}\approx1.316$ (for $\vartheta\to0$) to
$v_B^{\text{(nc)}}/v_E^{\text{(nc)}}\to(5/3)^{5/4}\approx1.894$ (for $\vartheta\to\infty$).

\subsubsection*{Information velocity}

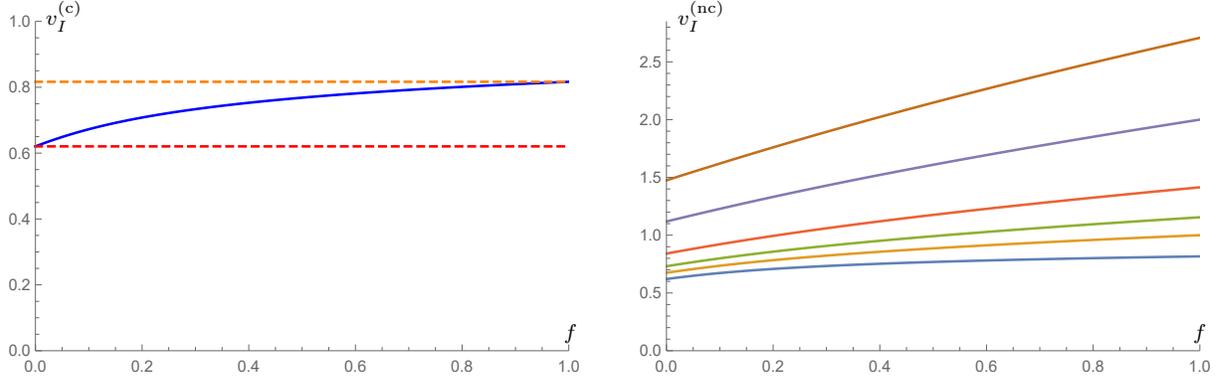
\begin{figure}[t!]
\centering
\includegraphics[width=3in]{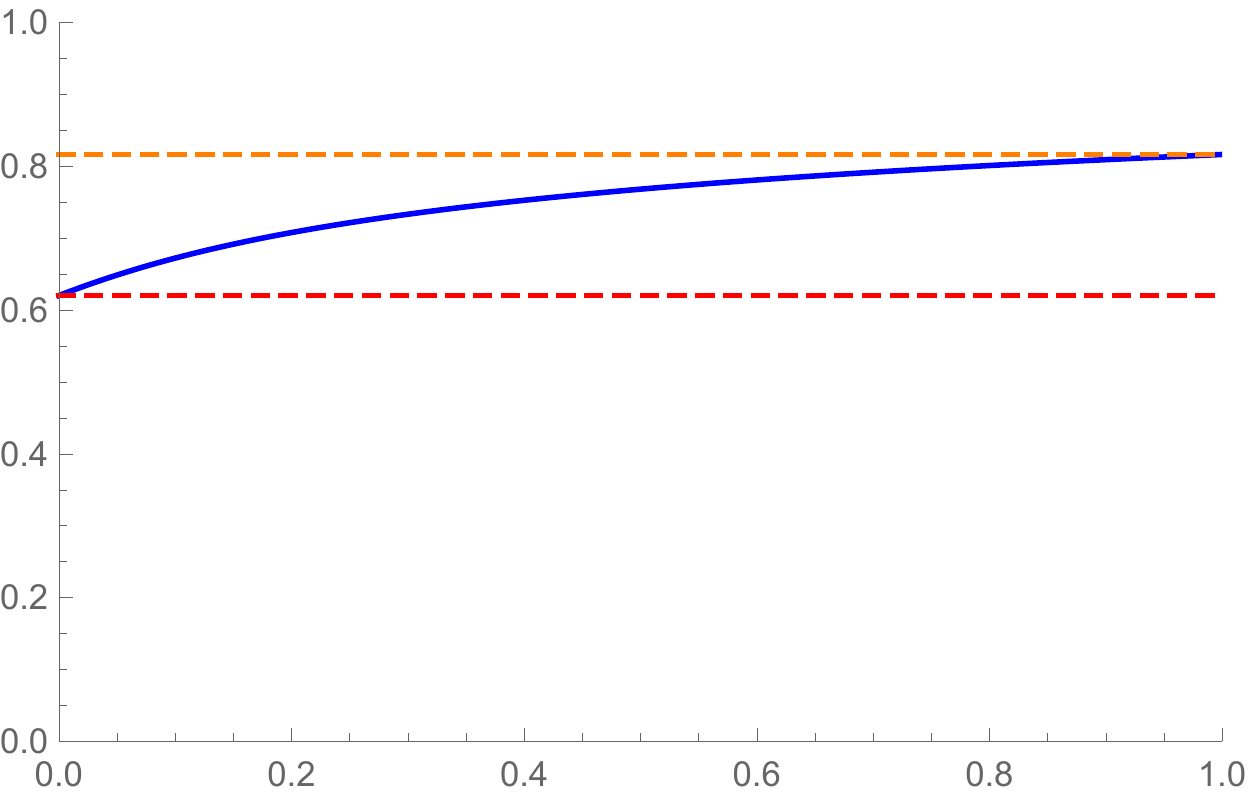}$\qquad$\includegraphics[width=3in]{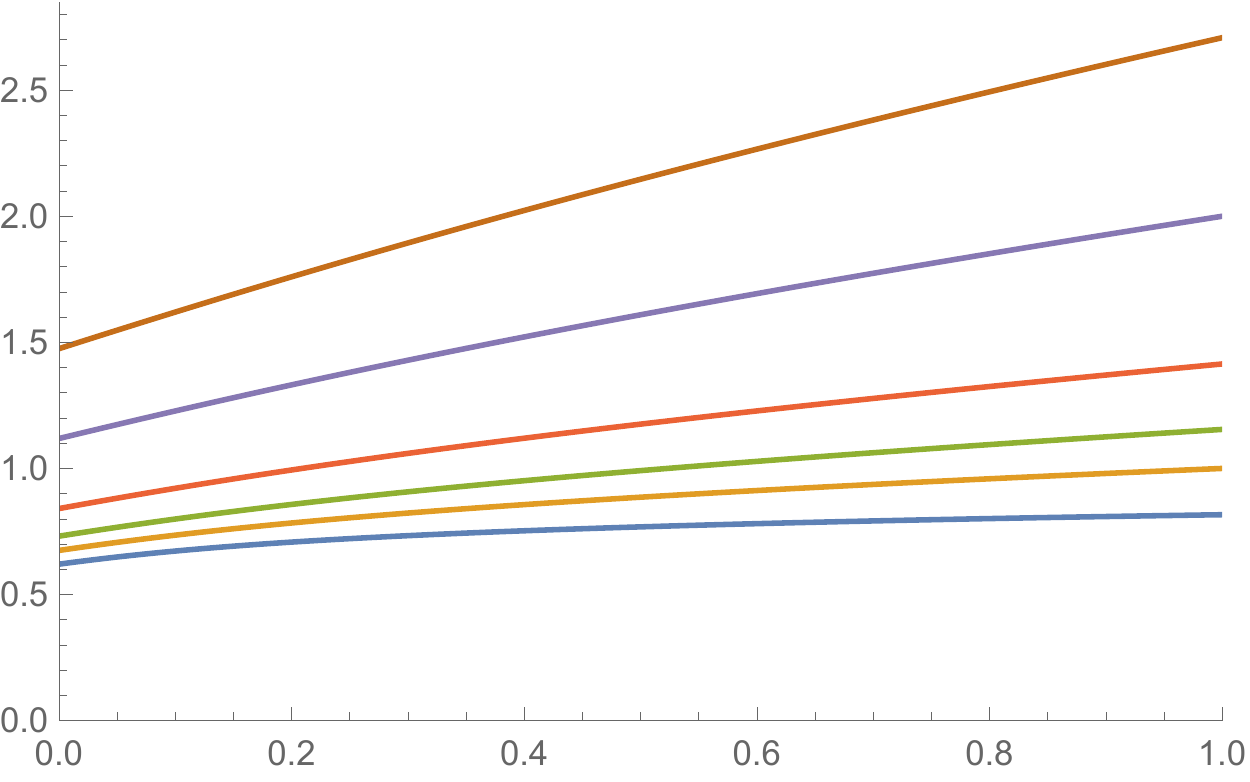}
  \begin{picture}(0,0)
\put(-207,149){{\scriptsize $v_I^{\text{(c)}}$}}
\put(32,149){{\scriptsize $v_I^{\text{(nc)}}$}}
\put(-11,30){{\scriptsize $f$}}
\put(227,30){{\scriptsize $f$}}
\end{picture}
 \caption{Left: Information velocity as a function of the entanglement fraction $f$ for the commutative strip. We also show the limiting values, $v_I^{\text{(c)}}(f=0)=v_E^{\text{(c)}}$ and $v_I^{\text{(c)}}(f=1)=v_B^{\text{(c)}}$, depicted in red and orange, respectively. Right: Information velocity for the non-commutative strip as a function of $f$ for various values of $\vartheta\in\{0,1/2,1,2,5,10\}$, from bottom to top, respectively. \label{plots2}}
\end{figure}

Finally, we can obtain the information velocity $v_I$ as a function of the entanglement fraction $f$ for . In order to do so, invert equation (\ref{fveq})
(to obtain $v_I(f)$) and then plug this into (\ref{v_Ifinal}). Unfortunately, it is not possible to invert (\ref{fveq}) analytically, even for the commutative strip. So we proceed numerically. In figure \ref{plots2} we show our findings for this quantity. In general we obtain that the $v_I$ increases monotonically both with $f$ and with $\vartheta$,
as opposed to the 1-D spin chain systems we studied previously. We believe that, in fact, large-$N$ is responsible for washing out the initial suppression of $v_I$ with respect to the nonlocal scale.

\subsubsection{Membrane theory for DDSYM}

We can analyze two cases, depending on the orientation of the strip. In both cases we find:
\be
P(\zeta)=\frac{R^{16}}{{\hat g}^{4}\zeta^6}\,,
\ee
however, the membrane tensions differ.
\begin{itemize}
  \item \textbf{Standard strip} ($\xi=\xi_1$ and $\vec{\xi}_\perp=\{\xi_2,\xi_3\}$)
\end{itemize}
This is completely analogous to the commutative strip. Here we repeat the analysis for completeness.
First, note that (\ref{EQMem}) can be written as:
\be
v^2=c_{\text{s}}(\zeta)\,,\qquad c_{\text{s}}(\zeta)=f(\zeta)-\frac{1}{6} \zeta f'(\zeta)=1-\frac{\zeta^4}{3 \zeta_h^4}\,,
\ee
and this equation can be inverted analytically to give:
\be
\zeta_\text{s}=c_{\text{s}}^{-1}(v^2)=\zeta_h\left[3 (1- v^2)\right]^{1/4}\,.
\ee
We have added a subindices ``s'' to indicate that we are working out the standard strip. Finally, the membrane tension is
\be\label{membraneS}
{\cal E}_{\text{s}}\left(v\right)=\zeta_h^3 \sqrt{-\frac{f'(\zeta)}{6 \zeta^5}}\Bigg\vert_{\zeta_\text{s}}=\sqrt{\frac{2}{3}}\frac{\zeta_h}{\zeta}\Bigg\vert_{\zeta_\text{s}}=\frac{\sqrt{2}}{3^{3/4} (1-v^2)^{1/4}}\,.
\ee
This is also the exact same result for a strip in Schwarzschild-AdS$_{5}$. It satisfies all the desired properties, including:
\be
 {\cal E}_{\text{s}}(0)=v_E \,, \qquad {\cal E}_{\text{s}}'(0)=0\,, \qquad {\cal E}_{\text{s}}(v_B)=v_B\,, \qquad {\cal E}_{\text{s}}'(v_B)=1\,,
\ee
with
\be
v_E^{\text{(s)}}=\frac{\sqrt{2}}{3^{3/4}}\,,\qquad v_B^{\text{(s)}}=\sqrt{\frac{2}{3}}\,.
\ee

\begin{figure}[t!]
\centering
\includegraphics[width=3in]{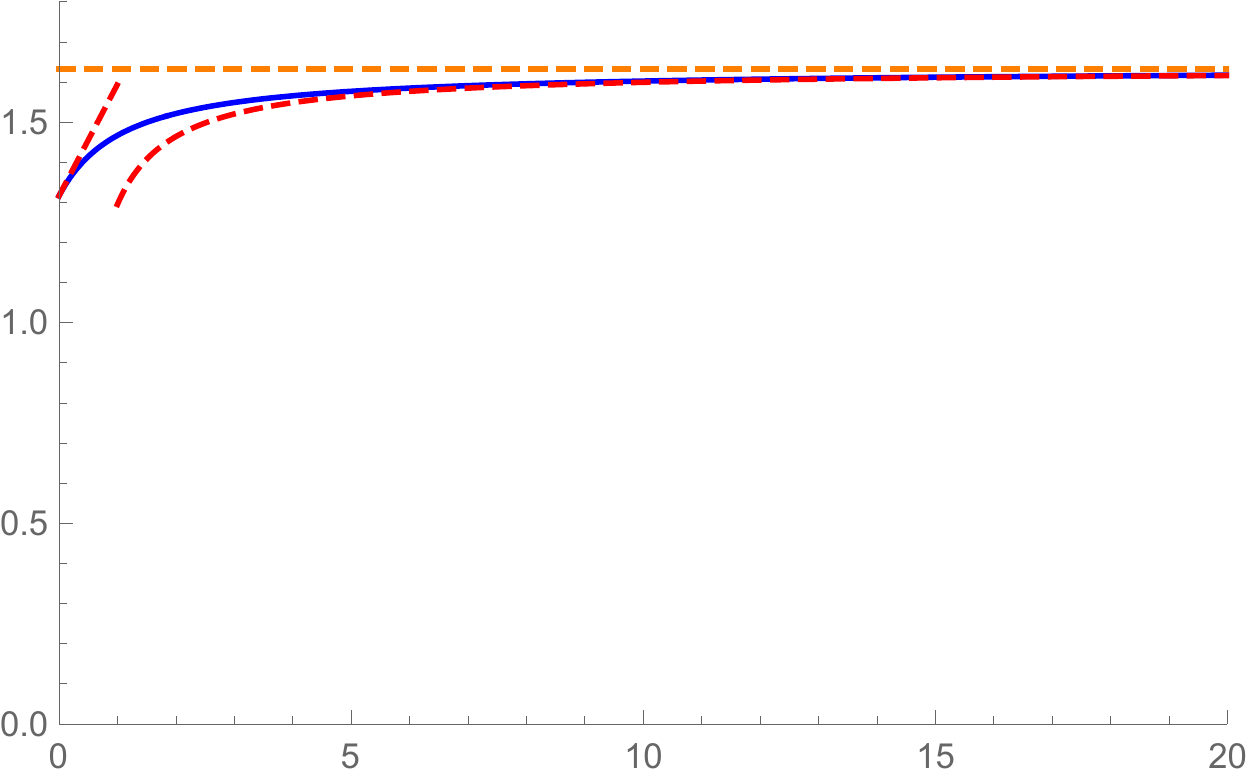}$\qquad$\includegraphics[width=3in]{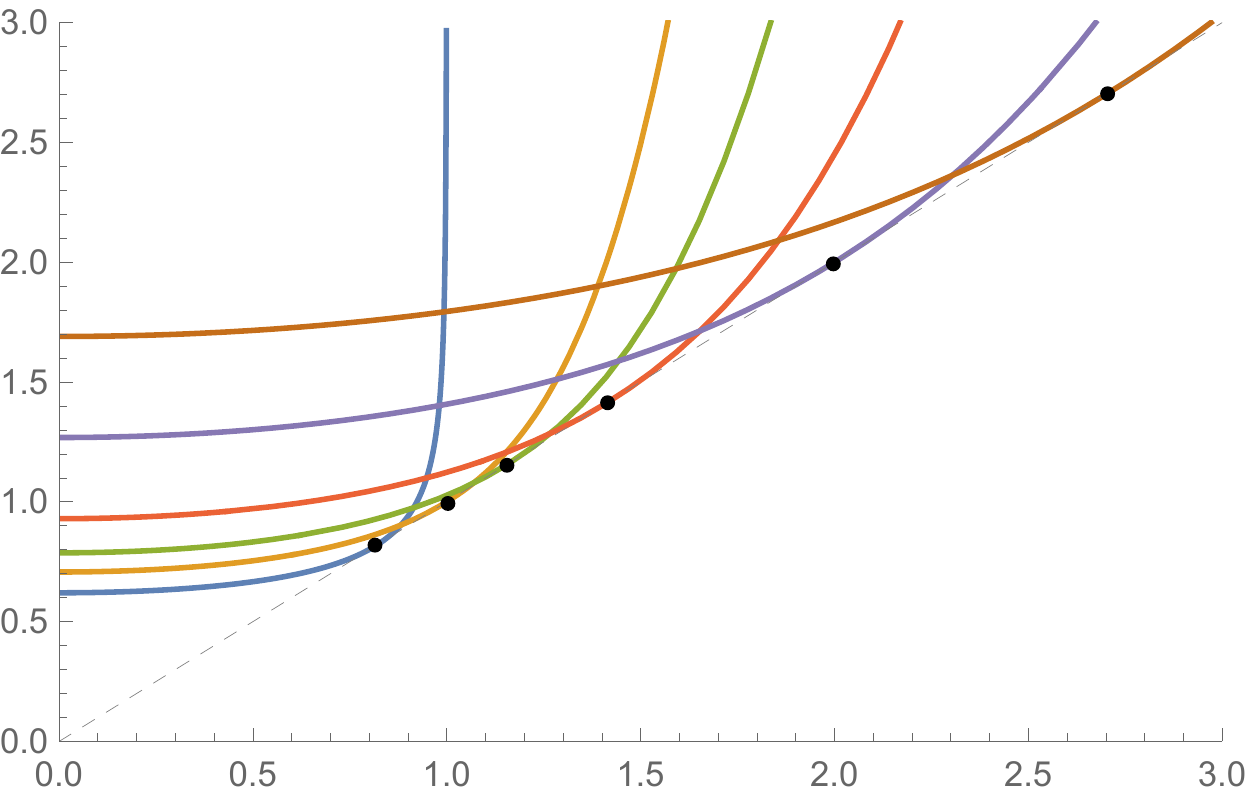}
  \begin{picture}(0,0)
\put(-207,149){{\scriptsize $v_B^{\text{(d)}}/v_E^{\text{(d)}}$}}
\put(32,149){{\scriptsize $\mathcal{E}_{\text{d}}(v)$}}
\put(-11,30){{\scriptsize $\delta$}}
\put(227,30){{\scriptsize $v$}}
\end{picture}
 \caption{Left: ratio between the butterfly velocity and the entanglement velocity as a function of the dimensionless combination $\delta=\lambda L^2T^2/4$. We also show the asymptotic value $v_B/v_E\to 2\sqrt{2/3}\approx1.633$, depicted in orange, and the expansions for $\delta\ll1$ and $\delta\gg1$, depicted in red. Right: Membrane tensions for $\delta\in\{0,1/2,1,2,5,10\}$, from bottom to top, respectively. In each case, we indicate the point at which the curve touches the straight line $\mathcal{E}(v_B)=v_B$.\label{plots1b}}
\end{figure}
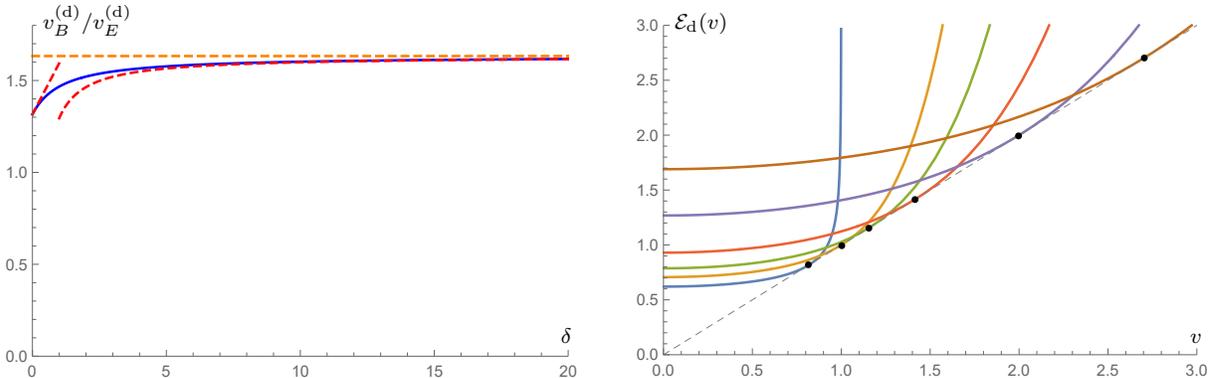

\begin{itemize}
  \item \textbf{Dipole oriented strip} ($\xi=\xi_3$ and $\vec{\xi}_\perp=\{\xi_1,\xi_2\}$)
\end{itemize}
In this case (\ref{EQMem}) can be written as:
\be\label{defcb}
v^2=c_{\text{d}}(\zeta)\,,\qquad c_{\text{d}}(\zeta)=\frac{f(\zeta)}{h(\zeta)}-\frac{\zeta f'(\zeta)}{6 h(\zeta)}+\frac{\zeta f(\zeta) h'(\zeta)}{6 h(\zeta)^2}=1-\frac{\zeta^4}{3 \zeta_h^4}+\frac{4 b^2}{3 \zeta^2}-\frac{2b^2z^2}{3\zeta_h^4}\,.
\ee
This expression can also be inverted analytically.\footnote{There are 6 roots (4 complex and 2 real), but only one coincides with the standard case in the limit $b\to0$.} However, the final expression is lengthy and not very illuminating so we will not transcribe it here. To have a flavor of the effects due to nonlocality, we expand the resulting functions in two limits. Defining a dimensionless parameter,
\be
\delta\equiv\frac{b^2}{\zeta_h^2}=\frac{\lambda L^2T^2}{4}\,,
\ee
we find that
\be
\zeta_\text{d}=c_{\text{d}}^{-1}(v^2)=\begin{cases}
  \displaystyle \zeta_h\left[3 (1- v^2)\right]^{1/4}\left[1-\frac{(1-3 v^2)\delta}{6 \sqrt{3} (1-v^2)^{3/2}}+\mathcal{O}\left(\delta^2\right)\right]& \displaystyle \qquad (\delta \ll 1)\,,\\[3ex]
  \displaystyle \zeta_h\, 2^{1/4}\left[1+\frac{(1-3 v^2)}{8 \sqrt{2} \delta}+\mathcal{O}\left(\frac{1}{\delta^2}\right)\right] & \displaystyle\qquad (\delta \gg 1)\,.
 \end{cases}
\ee
We have added a subindices `d'' to indicate that we are working out the dipole oriented strip. In this case, the membrane tension is
\bea
{\cal E}_{\text{d}}\left(v\right)&=&\zeta_h^3 \sqrt{-\frac{f'(\zeta)h(\zeta)-f(\zeta) h'(\zeta)}{6 \zeta^{5} h(\zeta)}}\Bigg\vert_{\zeta_{\text{d}}}=\sqrt{\frac{2}{3} \left[1+\frac{\zeta_h^2(\zeta^4+ \zeta_h^4)\delta}{2\zeta^6}\right]}\frac{\zeta_h}{\zeta}\Bigg\vert_{\zeta_{\text{d}}}\,,\nonumber\\
&=&\begin{cases}
  \displaystyle \frac{\sqrt{2}}{3^{3/4} (1-v^2)^{1/4}}\left[1+\frac{(2-3 v^2)\delta}{4 \sqrt{3}(1-v^2)^{3/2}}+\mathcal{O}(\delta^2)\right]& \displaystyle \qquad (\delta \ll 1)\,,\\[3ex]
  \displaystyle \frac{\sqrt{\delta}}{2}\left[1+\frac{(1+v^2)}{\sqrt{2}\delta}+\mathcal{O}\left(\frac{1}{\delta^2}\right)\right] & \displaystyle \qquad (\delta \gg 1)\,.
 \end{cases}
\eea
We can check that in the $\delta\to0$ limit we recover (\ref{membraneS}). At finite $\delta$ we can also verify (numerically, or analytically in some limits) that is satisfies all the desired properties:
\be
 {\cal E}_{\text{nc}}(0)=v_E \,, \qquad {\cal E}_{\text{nc}}'(0)=0\,, \qquad {\cal E}_{\text{nc}}(v_B)=v_B\,, \qquad {\cal E}_{\text{nc}}'(v_B)=1\,,
\ee
now with
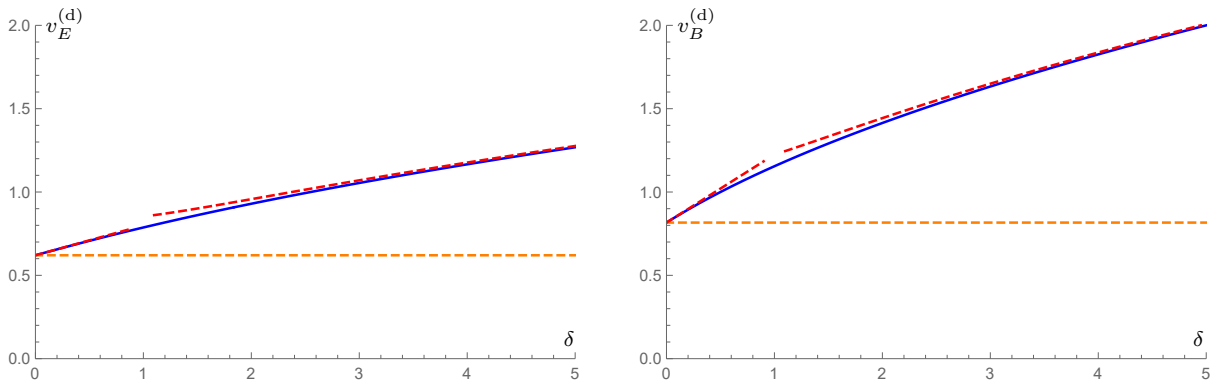
\begin{figure}[t!]
\centering
\includegraphics[width=3in]{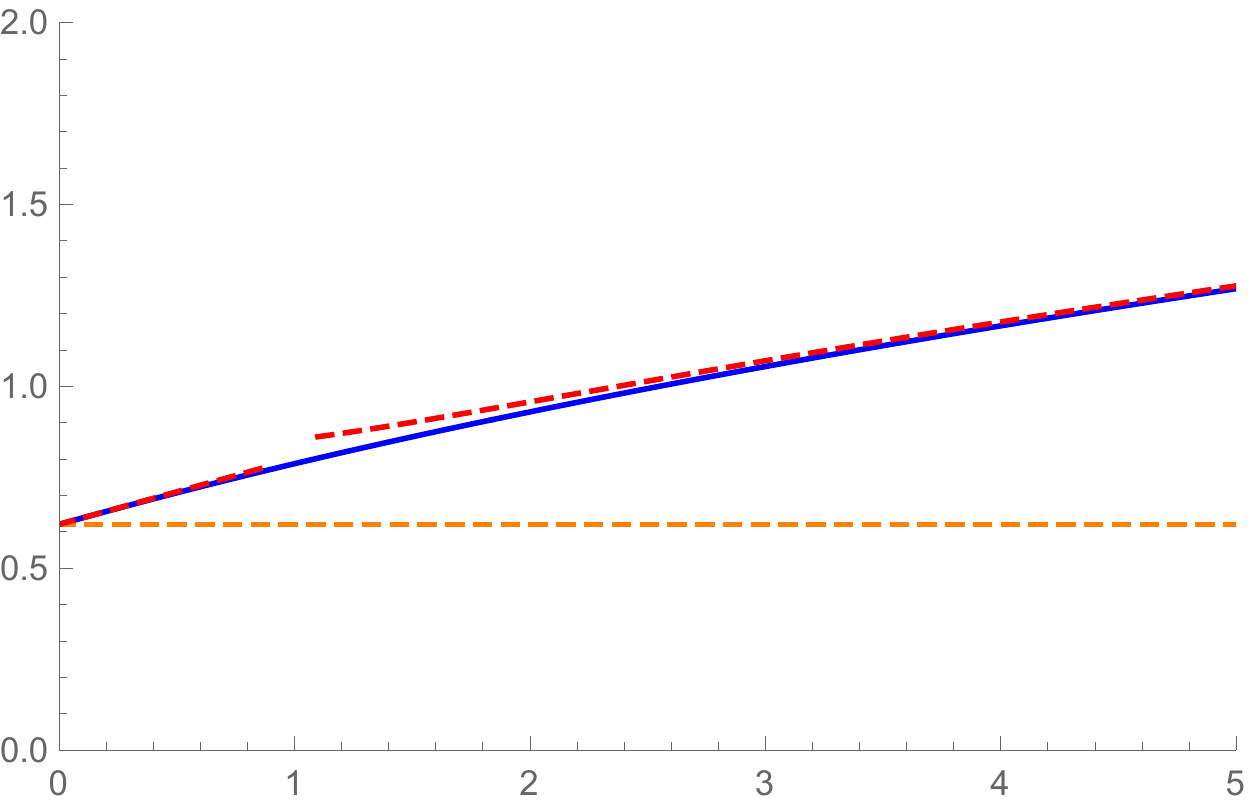}$\qquad$\includegraphics[width=3in]{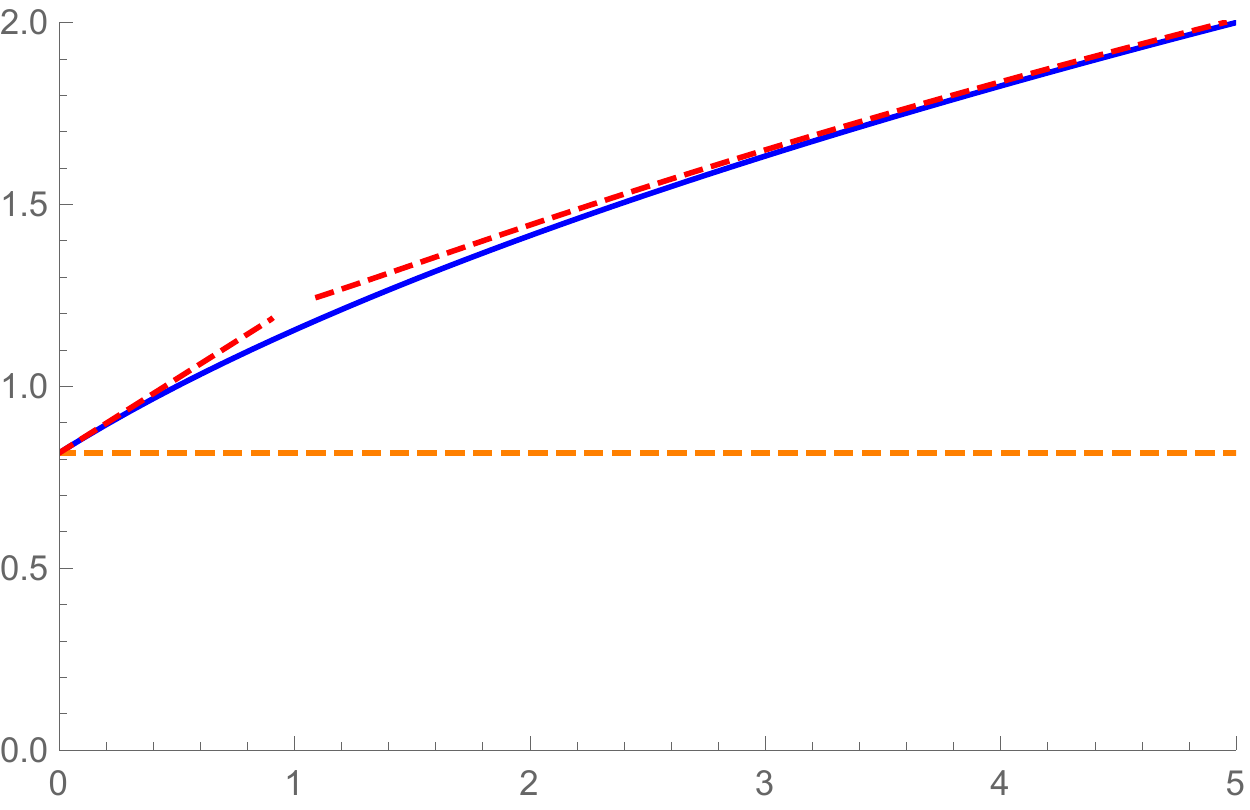}
  \begin{picture}(0,0)
\put(-207,149){{\scriptsize $v_E^{\text{(d)}}$}}
\put(32,149){{\scriptsize $v_B^{\text{(d)}}$}}
\put(-11,30){{\scriptsize $\delta$}}
\put(227,30){{\scriptsize $\delta$}}
\end{picture}
 \caption{Left: Entanglement velocity and Right: Butterfly velocity as a function of the dimensionless combination $\delta=\lambda L^2T^2/4$. In both plots we have shown in orange the standard values $v_E=\sqrt{2}/3^{3/4}\approx0.620$ and $v_B=\sqrt{2/3}\approx0.816$ obtained in the limit $\delta\to0$, and the expansions for $\delta\ll1$ and $\delta\gg1$, depicted in red. We notice that both velocities become superluminal at around $\delta\sim\mathcal{O}(1)$.\label{plotsvEBb}}
\end{figure}
\be\label{vEb}
v_E^{\text{(d)}}=\begin{cases}
  \displaystyle \frac{\sqrt{2}}{3^{3/4}}\left[1+\frac{\delta}{2\sqrt{3}}+\mathcal{O}(\delta^2)\right] & \displaystyle \qquad (\delta \ll 1)\,,\\[3ex]
  \displaystyle \frac{\sqrt{\delta}}{2}\left[1+\frac{1}{\sqrt{2}\delta}+\mathcal{O}\left(\frac{1}{\delta^2}\right)\right] & \displaystyle\qquad (\delta \gg 1)\,,
 \end{cases}
\ee
and
\be\label{vBb}
v_B^{\text{(d)}}=\begin{cases}
  \displaystyle \sqrt{\frac{2}{3}}\left[1+\frac{\delta}{2}+\mathcal{O}(\delta^2)\right] & \displaystyle \qquad (\delta \ll 1)\,,\\[3ex]
  \displaystyle \sqrt{\frac{2\delta}{3}}\left[1+\frac{1}{2\delta}+\mathcal{O}\left(\frac{1}{\delta^2}\right)\right] & \displaystyle\qquad (\delta \gg 1)\,.
 \end{cases}
\ee
In fact, we can obtain a very compact expression for $v_B$ by noticing that it can alternatively be obtained from (\ref{vBeasy}) and (\ref{defcb}):
\be
v_B=c(\zeta_h)^{1/2}=\sqrt{\frac{2}{3}(1+\delta)}\,.
\ee
Unfortunately, if we try to use (\ref{vEeasy}) to recover $v_E$ we run into similar problems as before, since the expression for the Hartman-Maldacena $\zeta_{\text{HM}}$ turns out to have a very complicated dependence with respect to $\delta$. We will therefore refrain from writing down a closed expression for $v_E$. Figure \ref{plotsvEBb} illustrate our main findings. In general, we notice that both (\ref{vEb}) and (\ref{vBb}) provide very good fits in their regime of applicability. We also observe that both velocities undergo a transition from subluminal to superluminal in the intermediate regime, where $\delta\sim\mathcal{O}(1)$ and ultimately diverge in the limit of very strong nonlocality $\delta\to\infty$. This is in qualitative agreement with the results we obtained for the NCSYM theory. Finally, we can also verify that $v_B^{\text{(d)}}>v_E^{\text{(d)}}$ for all $\delta$. Since both quantities diverge with the same power of $\delta$ ($v_B^{\text{(d)}}\propto \delta^{1/2}$, $v_E^{\text{(d)}}\propto \delta^{1/2}$) their ratio remains finite at large $\delta$. Expanding this ratio in the two limits we obtain:
\be
\frac{v_B^{\text{(d)}}}{v_E^{\text{(d)}}}=\begin{cases}
  \displaystyle 3^{1/4}\left[1+\frac{(3-\sqrt{3}) \delta}{6}+\mathcal{O}(\delta^2)\right] & \displaystyle \qquad (\delta \ll 1)\,,\\[3ex]
  \displaystyle 2 \sqrt{\frac{2}{3}}\left[1-\frac{(\sqrt{2}-1)}{2 \delta }+\mathcal{O}\left(\frac{1}{\delta^2}\right)\right] & \displaystyle\qquad (\delta \gg 1)\,.
 \end{cases}
\ee
These results are illustrated in figure \ref{plots1b}. From the plots we can confirm the monotonicity of $v_B^{\text{(d)}}/v_E^{\text{(d)}}$ with the strength of nonlocality. More specifically, the ratio interpolates between $v_B^{\text{(d)}}/v_E^{\text{(d)}}\to3^{1/4}\approx1.316$ (for $\delta\to0$) to
$v_B^{\text{(d)}}/v_E^{\text{(d)}}\to2\sqrt{2/3}\approx1.633$ (for $\delta\to\infty$). We see here that the nonlocal effects in the NC case are a bit more severe (in that case $v_B^{\text{(nc)}}/v_E^{\text{(nc)}}\to(5/3)^{5/4}\approx1.894$ as $\vartheta\to\infty$), which presumably can be traced back to the fact that in that case the nonlocality affects two of the spatial directions, while for the dipole theory only one of the directions.

\subsubsection*{Information velocity}

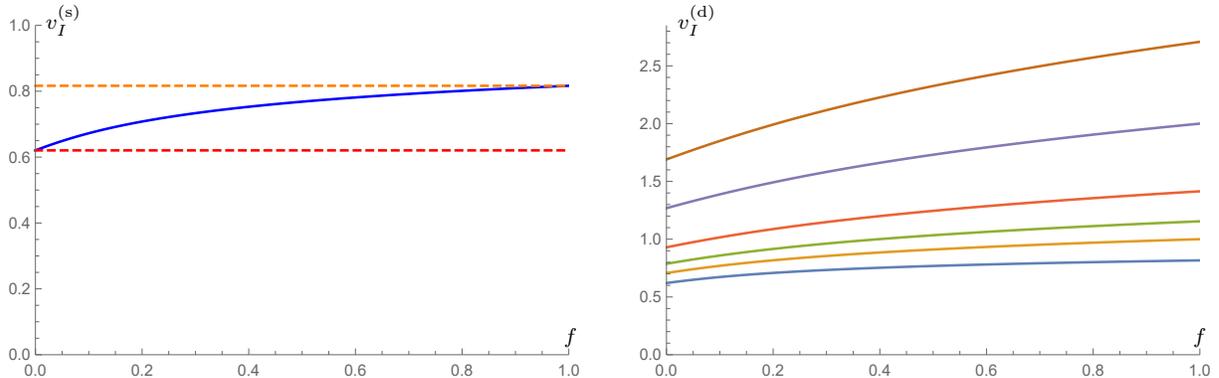
\begin{figure}[t!]
\centering
\includegraphics[width=3in]{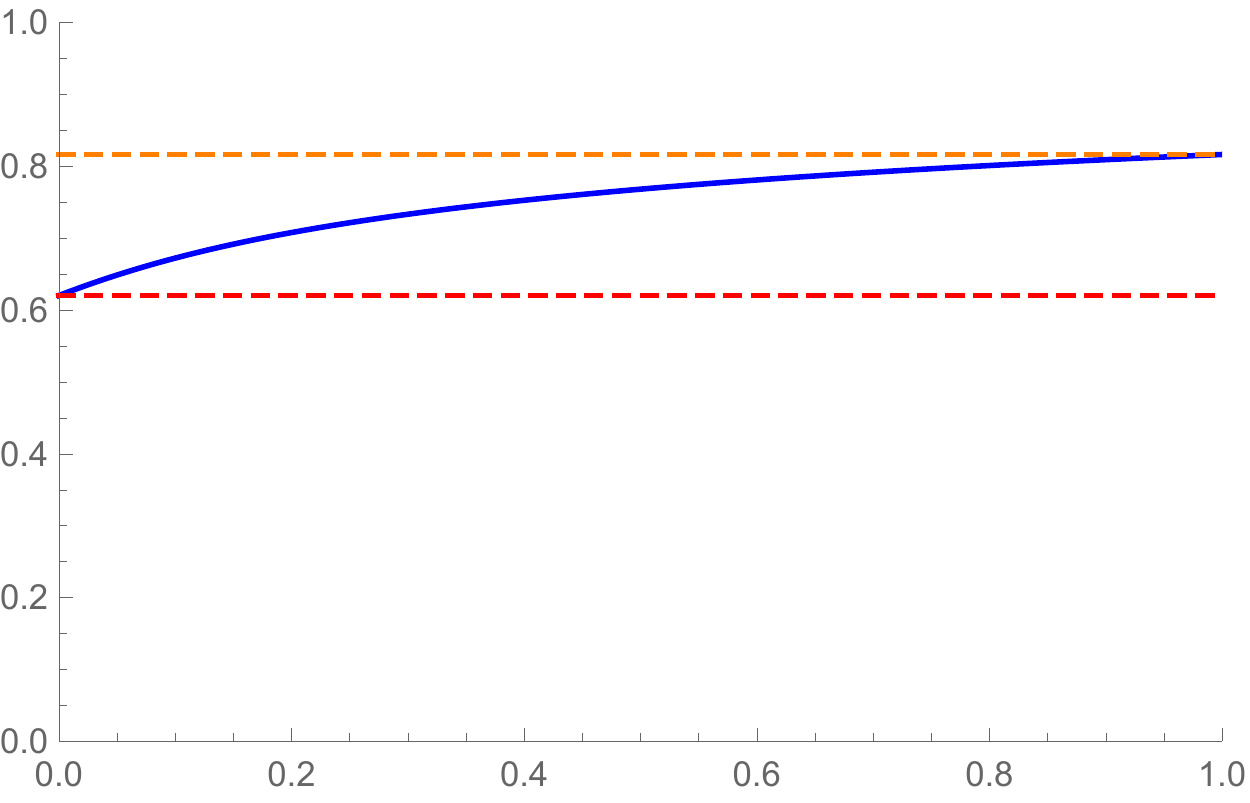}$\qquad$\includegraphics[width=3in]{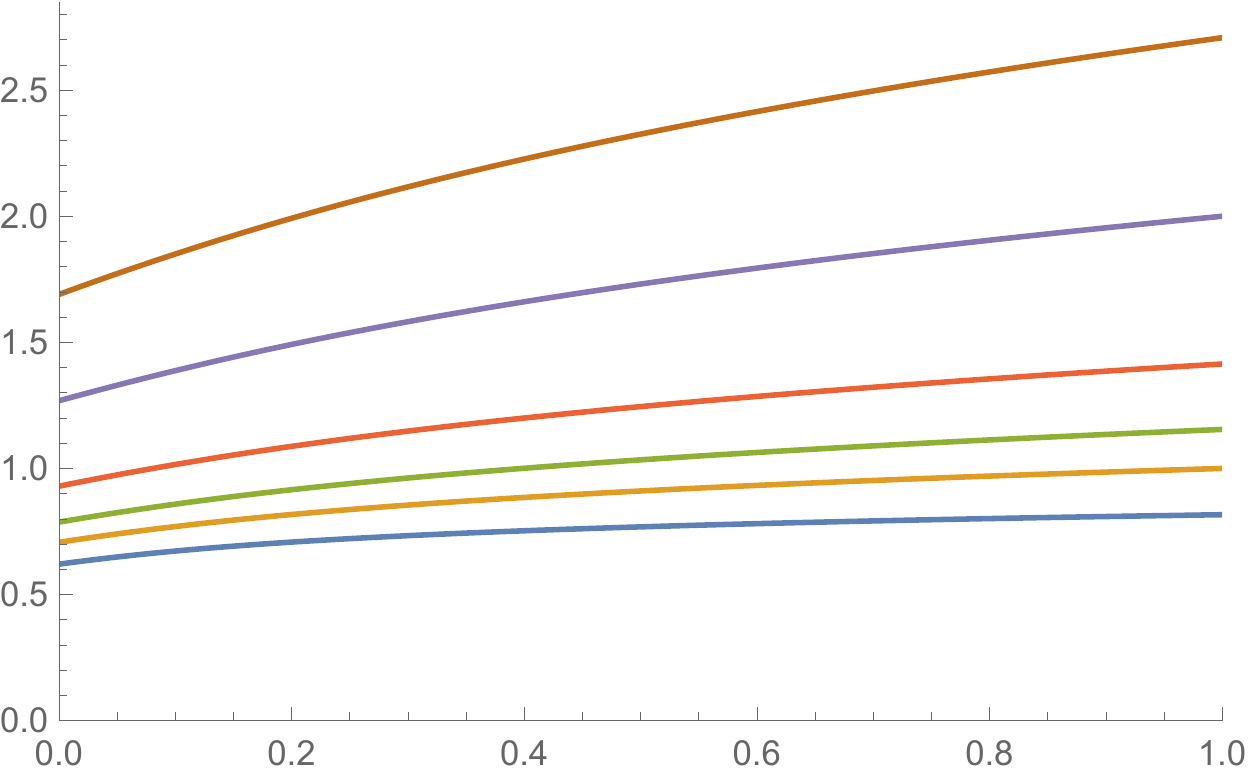}
  \begin{picture}(0,0)
\put(-207,149){{\scriptsize $v_I^{\text{(s)}}$}}
\put(32,149){{\scriptsize $v_I^{\text{(d)}}$}}
\put(-11,30){{\scriptsize $f$}}
\put(227,30){{\scriptsize $f$}}
\end{picture}
 \caption{Left: Information velocity as a function of the entanglement fraction $f$ for the standard strip. This reproduces the result for the commutative strip in figure \ref{plots2}. We also show the limiting values, $v_I^{\text{(s)}}(f=0)=v_E^{\text{(s)}}$ and $v_I^{\text{(s)}}(f=1)=v_B^{\text{(s)}}$, depicted in red and orange, respectively. Right: Information velocity for the dipole oriented strip as a function of $f$ for various values of $\delta\in\{0,1/2,1,2,5,10\}$, from bottom to top, respectively. This quantity behaves qualitatively the same as for the non-commutative strip, provided we identify the values of $\delta\leftrightarrow\vartheta$. However, there are very minor numerical differences for $f<1$ which make the curves for the dipole case slightly more concave. For $f=1$ they are exactly the same, because the functional dependence of $v_B$  with respect to $\delta$ or $\vartheta$ is the same for both theories. \label{plots2b}}
\end{figure}

Finally, we can obtain the information velocity $v_I$ as a function of the entanglement fraction $f$ for the DDSYM theory. In order to do so, invert equation (\ref{fveq})
(to obtain $v(f)$) and then plug this into (\ref{v_Ifinal}). Again, we proceed numerically here because it is not possible to invert (\ref{fveq}) even for the standard strip. In figure \ref{plots2b} we show our findings for this quantity. We obtain that the $v_I$ increases monotonically both with $f$ and with $\delta$, which was also found for the NCSYM system. This suggests that in large-$N$ systems the interference effects between the local and nonlocal couplings are effectively washed out.

\section{Conclusions and future work}
\label{sec:conclusions}

In this paper we have considered the possibility of speeding up the information transfer in a variety of systems by turning on mild nonlocalities. Our studies include 1-dimensional spin chains as well as strongly-coupled, large-$N$ theories with classical holographic duals. 
The nonlocal interactions that we consider only act below certain length scale and hence, and can be integrated out by a coarse groaning procedure. Thus, in the large-distance, late-time regime we do recover an effective notion of locality. Nevertheless, we find that these nonlocal interactions  induce an clear imprint in this regime, which in most cases translates to an enhancement in the rates of information transfer. Known systems that mimic this type of interactions include certain materials with high polarizability, which in the presence of a strong electromagnetic field undergo an ordered phase with their molecular dipoles aligned. We thus believe that our results could have practical applications in the areas of quantum computation and quantum communications.

The first part of the paper focused on the study of 1-dimensional spin chain systems with the addition of next-nearest neighbor and next-next-nearest neighbor interactions. We have computed the information velocity for states of uniform entanglement fraction, which confers not only the rate at which information propagates in these systems, but also the entanglement velocity $v_E$ and the butterfly velocity $v_B$, characterizing the rates of entanglement generation and operator growth, respectively.  Our results provide further support for the theory of information spreading of \cite{Couch:2019zni} and extend it to systems with mild nonlocalities.  We find that the addition of these extra couplings have a nontrivial effect on the information velocity, either suppressive (for very weak nonlocality), or enhancing $v_I$, though the generic behavior seems to be to raise $v_I$ (and $v_E$, $v_B$). The second part of the paper focused on holographic systems dual to strongly-coupled nonlocal gauge theories. In this case, we found that the suppressive behavior is completely washed out, in agreement with the earlier results in \cite{Fischler:2018kwt}. This behavior can possibly be explained as an effect of the large-$N$ limit. Indeed, preliminary results suggest that increasing the size of the local Hilbert space in the spin chain systems (e.g. considering qudits instead of qubits) seems to suppress the interference effects \cite{workinprogress}.

Some interesting directions for future work include:
\begin{enumerate}
\item 
Investigate the origin and nature of the interference effect found for some instances of small nonlocal couplings (see section \ref{subsec: next-nearest neighbor scans}) and understand how to maximize the information velocity $v_I$ from first principles.
\item
Explore the precise form of the information cone wavefront in the systems studied in this paper, along the lines of the $v_B$ wavefront analysis of \cite{Xu:2018dfp}.
\item
Test the effects of other forms of nonlocality chosen to match experimentally realizable setups, and explore more in detail possible practical applications.
\item
Investigate the effect of increasing the size of the local Hilbert space (e.g. qudits instead of qubits) to see what aspects of the phenomenology are brought closer to the large-$N$ limit represented in holographic systems \cite{workinprogress}.
\item
Study the universality of our holographic results (i.e. the suppression of the interference effects) in other nonlocal models. As an example, one may consider the dual to the near horizon limit of a stack of NS5-branes \cite{Aharony:1998ub}, also known as ``little string theory.''
\end{enumerate}
We hope to report back on some of these points in the near future.

\section*{Acknowledgements}
\label{sec:acknowledgements}

The authors would like to thank Josiah Couch and Phuc Nguyen for early collaboration on related topics.  We thank Mark Mezei for useful notes on employing the membrane method to compute $v_I$ in holography, and Shenglong Xu for helpful comments regarding spin chain computations.
We also acknowledge the Texas Advanced Computing Center (TACC) at The University of Texas at Austin for providing HPC resources that have contributed to the research results reported within this paper, \hyperlink{http://www.tacc.utexas.edu}{TACC}. The work of SE, WF, TG and SR is supported by the National Science Foundation under Grant Number PHY-1914679. JFP is supported by the Simons Foundation through \emph{It from Qubit: Simons Collaboration on Quantum Fields, Gravity, and Information}.

\providecommand{\href}[2]{#2}\begingroup\raggedright\endgroup

\end{document}